\documentclass{tlp}
\usepackage{aopmath}
\usepackage{graphicx}
\usepackage{fancyvrb}
\usepackage{algorithm}
\usepackage{algpseudocode}
\usepackage{pdfpages}

\usepackage[inline]{enumitem}
\usepackage[colorinlistoftodos,prependcaption,textsize=tiny]{todonotes}

\begin{document}
\bibliographystyle{acmtrans}

\long\def\comment#1{}

\title[Theory and Practice of Logic Programming]
        {Web-STAR: A Visual Web-Based IDE for a Story Comprehension System\thanks{An earlier version of this work was presented at the 2nd International Workshop on User-Oriented Logic Paradigms (IULP 2017). This article presents a newer version of the Web-STAR IDE with additional implemented features, along with the results of a user evaluation conducted to verify the usability, ease of use, and learnability of the IDE.}}

 \author[C. T. Rodosthenous and L. Michael]
         {CHRISTOS T. RODOSTHENOUS \and LOIZOS MICHAEL\\
         Open University of Cyprus\\
         \email{\{christos.rodosthenous,loizos\}@ouc.ac.cy}}

\pagerange{\pageref{firstpage}--\pageref{lastpage}}
\volume{\textbf{10} (3):}
\jdate{September 2017}
\setcounter{page}{1}
\pubyear{2017}

\maketitle

\label{firstpage}

\begin{abstract}
We present Web-STAR, an online platform for story understanding built on top of the STAR reasoning engine for STory comprehension through ARgumentation. The platform includes a web-based IDE, integration with the STAR system, and a web service infrastructure to support integration with other systems that rely on story understanding functionality to complete their tasks. The platform also delivers a number of ``social'' features, including a community repository for public story sharing with a built-in commenting system, and tools for collaborative story editing that can be used for team development projects and for educational purposes. Under consideration in Theory and Practice of Logic Programming (TPLP).
\end{abstract}


\section{Introduction}
\label{section:introduction}

Due to its long history, automated story understanding \cite{mueller:book:2006}, and text comprehension in general, has attracted interest from researchers across a diverse set of fields, including computer science, artificial intelligence, logic programming, psychology, language learning, narratology, and law. Researchers from these areas have varying interests in the different parts of the comprehension process, and varying skills in interacting with an automated comprehension system that seeks to implement this process. For story comprehension systems that rely on a symbolic representation of knowledge, for instance, one could roughly distinguish two groups of users: expert users (e.g., computer scientists, logic programmers, and AI experts), who might be more interested in developing systems for story understanding, and are able to encode and read stories and background knowledge in a machine-readable format; non-expert users (e.g., psychologists, language experts, narrators), who might be primarily interested in utilizing existing systems for story understanding, and may prefer to write stories in natural language, to examine the comprehension process and perform experiments, without caring that much about the internal encodings and representations that are used by the automated systems.

The distinction that we make between experts and non-experts is not meant to be absolute. Junior computer science students might fit better in the non-expert category, and language experts might be considered experts for the particular task of translating a story into a logic-based representation, even if they lack the skills to handle other parts of a story comprehension system. In any case, the diversity and heterogeneity that exists in terms of expertise in the use of automated story comprehension systems suggests the need for systems with a simple and intuitive interface that allows expert and non-expert users to input and reason with chosen stories, and to trace and debug the comprehension process. In this work, we start from an existing story comprehension system and build a platform on top of it to endow the end system with the aforementioned characteristics.

We consider, in particular, an existing automated story comprehension system called STAR: STory comprehension through ARgumentation \cite{Diakidoy2014,Diakidoy2015}, which adopts the view that comprehension requires the drawing of inferences about states and events that are not explicitly described in the story text \cite{Mueller:2003:SUT:1119239.1119246} through the use of background world knowledge and commonsense reasoning \cite{Mueller:2014:CRE:2821577}. Retaining the view that stories and background knowledge are symbolically represented, the STAR system abandons classical logic as the underlying semantics for knowledge, and adopts argumentation \cite{Bench-Capon2007,besnard2008elements} as a more appropriate substrate for the development of automated systems that interact with humans \cite{Kakas.etal_2016_ArgumentAndCognition,Michael_2017_TheAdviceTaker2.0}.

In this work we present the design and development of the \textbf{Web-STAR} platform built on top of the STAR system. The platform includes a web-based integrated development environment (IDE) that presents a personalized environment for each user with tools for writing, comprehending, and debugging stories, while visualizing the output of the comprehension process. Moreover, the IDE offers the basis for building a community, where people can share stories, comment, and reuse other community-created stories. Under the same umbrella, a web service is also made available for integrating other systems with the Web-STAR platform.

Web-STAR allows both expert and non-expert users to write stories and encode them in the internal STAR syntax, offering a number of features. Non-expert users can take advantage of the following features: \begin{enumerate*}[label=(\roman*)] \item the automatic conversion of a story from natural language to the STAR syntax, \item the encoding of background knowledge using a visual representation based on directed graphs, \item the automatic conversion of the graph to the STAR syntax and vice versa \end{enumerate*}. Non-expert users also benefit from the visual representation of the system output in a time-line format. Expert users benefit from the feature-rich IDE, which allows the preparation of a story in the STAR syntax using a state-of-the-art source code editor, the reasoner debugging options, and the raw output. All users benefit from the collaboration options available and the story repository.

In the following sections, we present the current state of affairs on story understanding systems and web-based IDEs that are used in logic-based systems, followed by a presentation of the STAR system as the underlying engine of the Web-STAR platform. Next, the Web-STAR platform is presented with details of the various features that it offers, along with scenarios on how these features can be used. The platform's usability is then evaluated and discussed, and new features and additions to the Web-STAR platform are presented as part of our ongoing work on the platform.

\section{Related work}
\label{section:related_work}
In this section, we provide insights on the concept of story understanding and its relation to the notion of explanation and we present related work on story understanding systems and on web-based IDEs that are geared towards imperative and declarative (logic-based) languages. Currently, little work has been done to enhance story understanding systems with functionality present in an IDE, and more specifically in a web-based IDE.

\subsection{Story understanding}
\label{subsection:story_understanding}
There are many definitions of what a story or a narrative is \cite{genette1984figures,prince2003dictionary,abbott2008cambridge,ricoeur2002narrative,ryan2007toward}, but in this context we use the view of \citeN{RickAltman2008}, as reported by \citeN{Michael2013}, who argues that ``virtually any situation can be invested with [those] characteristics [necessary to] perform the narrational function''. Story understanding includes the human ability to answer arbitrary questions, generate paraphrases and summaries, fill arbitrary templates, make inferences, reason about the story, hypothesize alternative versions of the story, look back over the story, and more \cite{Mueller99prospectsfor}.

Story understanding requires many different capabilities, including language understanding, word sense disambiguation, information extraction, along with the existence of vast amounts of background knowledge \cite{Davis:1990:RCK:83819,Lenat:1989:BLK:575523}. In the work of \citeN{Mueller99prospectsfor}, a historical overview of the story understanding area is presented along with the various challenges faced by a number of researchers. The difficulty of making progress on these challenges made a lot of researchers shift their attention to other areas that can bring more direct results.

\subsection{Story understanding systems}
\label{subsection:story_understanding_systems}
Several systems have been developed to date, to deal with the problem of story understanding and text comprehension. The majority of the developed systems are based on a symbolic representation of the story or script. These systems follow similar architectures in terms of how stories are encoded symbolically, how the background knowledge is encoded, and how the reasoning engine operates. There are also some attempts to employ different methods like deep learning \cite{NIPS2015_5782} and machine learning \cite{Ng:2000:MLA:1117794.1117810}. Story comprehension is typically tested through question answering, where a machine reads a passage of text and gives answers to questions. Through this test, the accuracy of the system is measured.

In the 1970's, Charniak presented two systems. In his early work \cite{Charniak1972}, a model was presented for answering questions related to children's stories by relating them to real-world background knowledge. The model used an internal representation language for the story and required an expert user to encode it. In his later work \cite{Charniak1973}, a program called Ms Malaprop was presented, for answering questions posed in a simple story. A semantic representation was used for encoding stories, questions, and answers in the program. The program used commonsense knowledge captured in the frame representation of ``mundane'' painting.

There is also work on Deep Read \cite{Hirschman1999}, an automated reading comprehension system that accepts stories and answers questions about them. Deep Read's creators used a corpus to conduct experiments using questions on stories with known answers. The system uses pattern matching techniques enhanced with automated linguistic processing. The system responds with a sentence that contains the correct answer in 30\%-40\% of the cases.

Another approach to story comprehension was Quarc \cite{Riloff:2000:RQA:1117595.1117598}, a rule-based system that reads a story and finds the sentence that best answers a given question. The system uses reading comprehension tests with questions on who, what, when, where, and why. Quarc (QUestion Answering for Reading Comprehension) uses lexical and semantic heuristics to look for evidence that a sentence contains the answer to a question. The rules used by the system are hand-crafted.

Similar to Deep Read is the work of \citeN{Wellner2006} on the ABC (Abduction-Based Comprehension) system. The system reads a text passage and answers test questions with short answer phrases as responses. It uses an abductive inference engine which allows first-order logical representation of relations between entities and events in the text and rules to perform inference over such relations. The system is also able to report on the types of inferences made while reasoning, allowing it to provide insights on where it is not performing well and give indications on where existing knowledge needs update or where new knowledge is required. The authors report an accuracy of 35\% using a strict evaluation metric.

More recent attempts include work by \citeN{mueller2007modelling} on a system which models space and time in narratives about restaurants. In particular, Mueller's system converts narrative texts into templates with information on the dialogues that take place in a restaurant, it uses these templates to construct commonsense reasoning problems, and then it uses commonsense reasoning and the created commonsense knowledge to build models of the dining episodes. Using these models, it generates questions and answers to the questions posed. The system was evaluated on stories retrieved from the Web and from the Project Gutenberg (https://www.gutenberg.org/). The evaluation showed that much work is needed for the system to produce highly accurate models.

A relatively new system is Genesis \cite{winston:OASIcs:2015:5290}, which deals with both story understanding and storytelling. It models and explores aspects of story understanding using stories drawn from sources ranging from fairy tales to Shakespeare's plays. The system uses the START parser \cite{Katz:1997:AWW:2856695.2856709} to translate English into a language of relations and events that the system can manipulate. This system is deployed using the Java WebStart mechanism.

Story understanding systems are not limited to text comprehension, but there is also work for other media like comic books, that combine textual content with pictures. In work by \citeN{DBLP:journals/corr/IyyerMGVBDD16}, deep neural architectures are tested on cloze-style tasks and the results show that text comprehension and image comprehension alone are not enough for a machine to comprehend a comic book story.

In terms of technical skills and expertise needed to use the aforementioned systems, the majority of them use a command line interface (CLI) and require users to prepare input files (e.g., story, background knowledge rules) using external tools. The output of these systems is generally in textual form, which makes it difficult for inspecting the resulting model. An exception to this is the Genesis system, which has a visual interface and provides a visual way to represent the output, but it is still a stand-alone application that requires installation on the user's device.

The lack of a visual online environment makes it harder for non-experts to setup and use these systems without prior programming knowledge and explicit knowledge of the specific system's internal mechanisms and representation. Furthermore, the majority of these systems rely on external tools (text editors) for editing the source code and lack basic functionality that an IDE can easily provide (e.g., code folding, syntax highlighting).

\subsection{Narratives and explanation}
\label{subsection:story_understanding_explanation}
The notion of story understanding is closely related to the notion of explanation. There has been an interesting debate and work on this connection, mostly from the philosophical point of view. \citeN{10.2307/3595560} state that ``A story does more than recount events; it recounts events in a way that renders them intelligible, thus conveying not just information, but also understanding. We might therefore be tempted to describe narrative as a genre of explanation.''. \citeN{carroll2001narrative} describes a narrative as a common form of explanation since it is usual to use narratives to explain how things happened. This is also connected to the causal relations of the events in a narrative. \citeN{forster2010aspects} uses the term ``plot'' to describe a story that is distinguished by the ``why?'' question and to separate it from one that is connected with the ``and then?'' question. The first is a form of explanation since one needs to answer the ``why'' question that includes a causal link between the story concepts. The work of \citeN{10.2307/40970551} includes a discussion on whether narratives explain.

Other work concentrates on explanation and scientific understanding \cite{10.2307/2024924} and tries to combine the use of narratives with explanation of scientific notions with the purpose to answer why and how questions. Recent work by \citeN{Morgan2017} investigates the role of narratives in the social science case-based research, by creating a productive ordering of the materials within such cases, and on how such ordering functions in relation to ``narrative explanation''.

The Web-STAR IDE allows the encoding of both causal rules in the background knowledge and questions in the story that provide explanations on the story concepts. Moreover, the debugging options offered by the Web-STAR IDE provide in-depth insights on the inferences made to provide answers to questions and hence lead to the relevant explanation.

\subsection{Web-based IDEs}
\label{subsection:web_ides}
Web-based IDEs are systems available through a web browser with no reliance on specific hardware or software stack and are agnostic to the Operating System. Some of these are now considered as mainstream IDEs for developing applications, such as Cloud9 \cite{cloud9website}, Codiad \cite{codiadwebsite}, ICEcoder \cite{icecoderwebsite}, codeanywhere \cite{codeanywherewebpage} and Eclipse Che \cite{eclipsechewebsite}.

These web-based IDEs allow users to write code in an online source code editor using the programming language of their choice. They also provide code folding, code highlighting, and auto-complete functionality, built in their source code editors. Moreover, some of them provide online code execution functionality with access to an underlying virtual machine and thus access to the shell.

Web-based IDEs are also used in the logic programming domain. There are only a few systems developed to address this need, like SWISH (SWI-Prolog for Sharing) \cite{Wielemaker2015}, IDP Web-IDE \cite{Dasseville2015}, and Answer Set Programming (ASP) specific IDEs and tools, like the system presented in the work of \citeN{DBLP:journals/corr/MarcopoulosRZ17}.

SWISH is a web front-end for SWI-Prolog, and is used to run small Prolog programs for demonstration, experimentation, and education. The platform offers collaborative tools for users to share programs with others, and a chat functionality. An instantiation of the system was also used to build the SWISH DataLab system \cite{Bogaard2017}, which is oriented towards data analysis.

Additionally, there is also a web-based implementation of the computer language LPS (Logic-based Production System) \cite{Kowalski2016} built as an extension of SWISH. This system aims at supporting the teaching of computing and logic in secondary school.

The SWISH design is geared towards the educational domain, allowing learners of the Prolog language to easily access code examples and execute them without the need to install SWI-Prolog locally. It exposes only a limited subset of the SWI-Prolog language, and it is not recommended for large and real-world applications.

The IDP Web-IDE is an online front-end for Imperative Declarative Programming (IDP), a Knowledge Base System for the FO($\cdot$) language. FO($\cdot$) is an extension of first-order logic (FO) with types, aggregates, inductive definitions, bounded arithmetic and partial functions \cite{Denecker:2008:LNI:1342991.1342998}. The Web-IDE allows users to open a chapter from the online tutorial and start testing example programs. There are options for collaborative work and visualization functionality for some of the program outputs.

In the work of \citeANP{DBLP:journals/corr/MarcopoulosRZ17}, an online system is presented with a cloud file system and a simple interface, which allows users to write logic programs in the SPARC language \cite{Balai2013} and perform several tasks over the programs. The authors aim to use this system to teach Answer Set Programming to undergraduate university students and high school students.

\section{The STAR system}
\label{section:star_system}
Based on the successful use of web-based IDEs in logic-based systems, we present in this work the development of a web-based IDE for the STAR system. The STAR system is based on the well-established argumentation theory in Artificial Intelligence \cite{Baroni:2011:RIA:2139707.2139708,Bench-Capon2007}, uniformly applied to reason about actions and change in the presence of default background knowledge \cite{Diakidoy2015}. The STAR system follows guidelines from the psychology of comprehension, both for its representation language and for its computational mechanisms for building and revising a comprehension model as the story unfolds.

In terms of its underlying infrastructure, the STAR system is written in SWI-Prolog \cite{wielemaker_schrijvers_triska_lager_2012}. Upon the setting up of the Prolog environment, and the invocation of the system, a user-selected domain file is loaded and processed. We present the syntax and semantics of the STAR system through the example story in Fig.~\ref{fig:example:story_nl}.
\begin{figure}
\figrule
\begin{center}
\begin{Verbatim}[numbers=left,xleftmargin=5mm]
Bob called Mary on the phone.

Was Mary embarrassed?
Was the phone ringing?

She did not want to answer the phone.
Bob had asked her for a favor.
She had agreed to do the favor.

Was the phone ringing?

She answered the phone.
She apologized to Bob.

Was Mary embarrassed?
\end{Verbatim}
\end{center}
\caption{A short story in natural language with interspersed questions.}
\label{fig:example:story_nl}
\figrule
\end{figure}

The example story is interspersed with questions. These are not meant to be parts of the story, but are questions directed towards the reader of the story. Whenever a sequence of questions is encountered, the reader is expected to provide answers to the questions based on the information given in the story in all the preceding story lines. The story then continues until a new sequence of questions is encountered, and so on. Each such part of the story is effectively a \emph{scene} or a reading \emph{session}, and each session is associated with the questions that need to be answered based on the information provided in that and all preceding story sessions. Although the reader goes through the story in the linear fashion in which the story is represented, the story time need not be linear, and can jump back and forth between different time periods. Questions are assumed to refer to a story time-point following the one at which the last story session left off.

As the story unfolds, answers to questions might change either because the same question is asked at a different point in the story time-line, or because the story information leads the reader to revise their \emph{comprehension model} of what they infer (based on their background knowledge and the given story information) to be the case in the story world. The two questions in the example have their answers changed as a reader progresses from the top to the bottom of the story, with the question ``Was the phone ringing?'' changing because of the first of the aforementioned reasons, and the question ``Was Mary embarrassed?'' changing because of the second of the aforementioned reasons.

Having explained how a reader may comprehend our example story, we hasten to note that the STAR system \emph{does not} process stories in natural language --- in fact, processing stories in natural language is one of the main features of the web-based IDE that is presented in this work. Instead, the STAR system expects the story statements, their partitioning into sessions, and their association with questions to be provided in a certain symbolic language. All these elements constitute the first part of the domain file that the STAR system loads once it is invoked. A possible representation (although by no means the only one) of our example story is given in Fig.~\ref{fig:example:story_star}.
\begin{figure}
\figrule
\begin{center}
\begin{Verbatim}[numbers=left,xleftmargin=5mm]
session(s(0),[],all).
session(s(1),[q(1),q(2)],all).
session(s(2),[q(3)],all).
session(s(3),[q(4)],all).

s(0) :: is_favor(favor1) at always.
s(0) :: is_person(bob) at always.
s(0) :: is_person(mary) at always.
s(0) :: is_phone(phone1) at always.

s(1) :: call(bob, mary, phone1) at 6.
s(2) :: -do_want(mary,answer(phone1)) at 12.
s(2) :: have_ask(bob, mary, favor1) at 2.
s(2) :: have_agreed(mary,do(favor1)) at 4.
s(3) :: answer(mary, phone1) at 16.
s(3) :: apologize(mary, bob) at 18.

q(1) ?? is_embarrassed(mary) at 8.
q(2) ?? is_ringing(phone1) at 10.
q(3) ?? is_ringing(phone1) at 14.
q(4) ?? is_embarrassed(mary) at 20.
\end{Verbatim}
\end{center}
\caption{A possible representation of the example story depicted in Fig.~\ref{fig:example:story_nl} to the STAR syntax.}
\label{fig:example:story_star}
\figrule
\end{figure}

Each of the story statements is of the form \texttt{s(N) $::$ Literal at Time-Point}, where \texttt{N} is a non-negative integer representing the session of that statement; session \texttt{0} is a special session that includes typing information only. A literal \texttt{Literal} is either a concept \texttt{Concept} or its negation -\texttt{Concept} (i.e., the symbol for negation is ``-''), where a concept \texttt{Concept} is a predicate name along with associated variables or constants for the predicate's arguments. The representation of our example story clearly shows its non-linear time-line.

Following the story statements are the question statements of the form \texttt{q(N) $??$ Literal at Time-Point; Literal at Time-Point; \dots.}, where \texttt{N} is a non-negative integer representing the number of the question and ``;'' separates the possible answers to that question; although the notation is meant to represent multiple-choice questions, in effect the STAR system treats each of the choices as a true/false question. Which questions are associated with which sessions is given by the session statements.

Given the story and question representation in Fig.~\ref{fig:example:story_star}, the STAR system aims to produce a comprehension model of the story, through which it will subsequently attempt to answer the posed questions. Much like human readers, the STAR system invokes background knowledge about the story world to infer what else holds beyond what is explicitly stated in the story. This background knowledge is also represented in a logic-based language, and constitutes the second part of the domain file. For our example story, and in a manner consistent with our chosen symbolic representation of that story, a possible representation of (some of) the background knowledge relevant for the story is given in Fig.~\ref{fig:example:bg_story_star}.

\begin{figure}
\figrule
\begin{center}
\begin{Verbatim}[numbers=left,xleftmargin=5mm]
fluents([
   do_want(_,_),
   is_embarrassed(_),
   carried_out(_),
   has_asked_for(_,_,_),
   has_agreed_to(_,_),
   is_ringing(_)
]).

c(01) :: have_ask(P1,P2,S) causes has_asked_for(P1,P2,S).
c(02) :: have_agreed(P2,do(S)) causes has_agreed_to(P2,S).

p(11) :: has_asked_for(P1,P2,S), has_agreed_to(P2,S), apologize(P2,P1)
         implies -carried_out(S).

c(21) :: have_agreed(P2,do(S)), -carried_out(S) causes is_embarrassed(P2).

c(31) :: has_asked_for(P1,P2,S), has_agreed_to(P2,S), -carried_out(S),
         call(P1,P2,D), is_phone(D) causes -do_want(P2,answer(D)).

c(41) :: is_person(P1),is_person(P2),call(P1,P2,D),is_phone(D) causes
         is_ringing(D).

c(42) :: is_person(P1),answer(P1,D),is_phone(D) causes -is_ringing(D).

c(42) >> c(41).
\end{Verbatim}
\end{center}
\caption{A possible representation of the background knowledge for comprehending the story depicted in Fig.~\ref{fig:example:story_nl}, assuming its encoding in Fig.~\ref{fig:example:story_star}}.
\label{fig:example:bg_story_star}
\figrule
\end{figure}

The presented representation includes four type of statements\footnote{The STAR syntax allows additional types of statements and expressivity, which we do not present here for simplicity.}: a list of concepts that are marked as \emph{fluents}, indicating that their truth value persists across the story time-line; rules prefixed by the symbols \texttt{c(N)} and \texttt{p(N)} to indicate that they are, respectively, causal or property rules, and priorities \texttt{>>} indicating relative strength between conflicting rules.

The main part of a rule is of the form \texttt{Body causes / implies Head.}, where \texttt{Body} is either the tautology \texttt{true}, or a comma-separated list of literals, and \texttt{Head} is a single literal. Each rule, then, expresses an implication from the premises in its body to the conclusion in its head. The difference between the causal and property rules lies in their treatment of time. Property rules are meant to capture dependencies between the properties of entities, and refer to any single point in the story time-line: whenever the body holds, the head also holds at that same time-point. On the other hand, causal rules are meant to express how things change over time, and capture dependencies between consecutive time points: whenever the body holds, the head holds at the following time-point. Further, when the head literal of a causal rule is inferred, this inference \emph{causes} the persistence of the truth-value of that literal from earlier time points to stop in case the persisted truth-value conflicts with the inference. In effect, the fluent list expresses implicitly a third type of persistence rules, along with the implicit lower priority compared to all conflicting causal rules. Additional priorities between causal and/or property rules are expressed explicitly.

We will not discuss in detail the intuitive interpretation of the rules in the background knowledge for our example story, other than to say that they roughly capture the knowledge that: if you apologize to someone that has asked you to do something, to which you have agreed, then it is because you have not carried out that something; having agreed to do something that you have not carried out causes embarrassment, and further causes not wanting to answer the phone when the call is from the person that has asked you for that something; a call causes the phone to start ringing, and answering the phone causes the ringing to stop.

With all the aforementioned logic-based information in a domain file, the STAR system proceeds to construct a comprehension model of the story. A comprehension model can be thought as a partial mapping from timed concepts to truth-values, essentially indicating when each concept is true, false, or unknown. In computing these truth-values, one takes into account both the information given explicitly in the story, but also draws inferences through the background knowledge. The STAR system adopts a particular argumentation-based approach to how inferences are drawn. Roughly, it combines story statements with rules to build a proof of the entailment of a literal. Rules are used both in the forward direction (i.e., via modus ponens) and in the backward direction (i.e., via modus tolens). Since different combinations of story statements and rules might lead to contradictory inferences, each constructed proof is viewed as an argument in support of an inference, and conflicts between arguments are resolved by lifting the priority relation between rules to an attacking relation between arguments \cite{Rahwan:2009:AAI:1594368}.

Without discussing the technical nuances of the particular argumentation approach that the STAR system adopts \cite{Diakidoy2014}, we briefly comment on how the STAR semantics relate to existing well-established argumentation semantics. STAR adopts a structured rule-based argumentation framework in the spirit of ASPIC+ \cite{doi:10.1080/19462166.2013.869766}. Combinations of premises from the story with defeasible rules from the background knowledge form a proof tree in support of some inference; this tree corresponds to an argument. Unlike in the ABA framework \cite{doi:10.1080/19462166.2013.869878} the premises are assumed to be indefeasible, and (exogenously) attack any argument that supports a contrary inference. Also unlike in the ABA framework, arguments (endogenously) attack each other on the rules they use, not on their premises. An attack comes from the last / head / top rule in the proof tree of an argument, and is directed towards any (possibly internal) rule in the proof tree of another argument. As long as the former rule is not less preferred than the latter rule, the attack is present. The semantics of the attack relation implies, in particular, that a pair of arguments can attack each other. With this attack relation, the STAR system proceeds to compute the grounded extension of the resulting argumentation framework, and offers this unique extension as the comprehension model of the story. Beyond consulting the relevant work for more details \cite{Diakidoy2014}, the interested reader may wish to also consult a more recent work \cite{Michael_2017_TheAdviceTaker2.0}, where a similar in spirit argumentation semantics is discussed, without the nuances of temporal reasoning and contrapositive reasoning, and where a case is also made for the learnability of this type of arguments.

Once the grounded extension is computed after each session (with the arguments that are relevant given the premises of the story that far), the STAR system outputs the computed comprehension model, as in Fig.~\ref{fig:example:star_story_output} for our example story.

\begin{figure}
\figrule
\begin{center}
\begin{Verbatim}[numbers=left,xleftmargin=5mm]
===================================
>>> Reading story up to scene s(3)
===================================
>>> Universal argument...
>>> Acceptable argument...

>>> Comprehension model:

0: -carried_out(favor1) < is_favor(favor1)> < is_person(bob)>
   < is_person(mary)> < is_phone(phone1)> < is_ringing(ringing1)>
   < is_ringing(ringing2)>

1: -carried_out(favor1) < is_favor(favor1)> < is_person(bob)>
   < is_person(mary)> < is_phone(phone1)> < is_ringing(ringing1)>
   < is_ringing(ringing2)>

2: -carried_out(favor1) < is_favor(favor1)> < is_person(bob)>
   < is_person(mary)> < is_phone(phone1)> < is_ringing(ringing1)>
   < is_ringing(ringing2)> < have_ask(bob,mary,favor1)>

...

18: -carried_out(favor1) is_embarrassed(mary) < is_favor(favor1)>
    < is_person(bob)> < is_person(mary)> < is_phone(phone1)>
    -is_ringing(phone1) < is_ringing(ringing1)> < is_ringing(ringing2)>
    < apologize(mary,bob)> -do_want(mary,answer(phone1))
    has_agreed_to(mary,favor1) -call(bob,bob,phone1)
    -call(bob,mary,phone1) -call(mary,bob,phone1) -call(mary,mary,phone1)
    has_asked_for(bob,mary,favor1)

19: -carried_out(favor1) is_embarrassed(mary) < is_favor(favor1)>
    < is_person(bob)> < is_person(mary)> < is_phone(phone1)>
    -is_ringing(phone1) < is_ringing(ringing1)> < is_ringing(ringing2)>
    -do_want(mary,answer(phone1)) has_agreed_to(mary,favor1)
    -call(bob,bob,phone1) -call(bob,mary,phone1) -call(mary,bob,phone1)
    -call(mary,mary,phone1) has_asked_for(bob,mary,favor1)

20: -carried_out(favor1) is_embarrassed(mary) < is_favor(favor1)>
    < is_person(bob)> < is_person(mary)> < is_phone(phone1)>
    -is_ringing(phone1) < is_ringing(ringing1)> < is_ringing(ringing2)>
    -do_want(mary,answer(phone1)) has_agreed_to(mary,favor1)
    has_asked_for(bob,mary,favor1)

>>> Answering question q(4):
+ accepted choice: ,[is_embarrassed(mary)at 20]

>>> Finished reading the story!
\end{Verbatim}
\end{center}
\caption{Part of the output of the STAR system for the story depicted in Fig.~\ref{fig:example:story_nl}, as encoded in Fig.~\ref{fig:example:story_star} and with the associated background knowledge presented in Fig.~\ref{fig:example:bg_story_star}.}
\label{fig:example:star_story_output}
\figrule
\end{figure}

The output presents the comprehension model (literals that are true at each time-point), with parts of it marked in triangular parentheses to indicate that those come directly from the story and not from inferences. Following the comprehension model, each of the questions (of the session being processed) are presented along with all their choices for answers, and for each answer the system responds on whether it is accepted, rejected, or possible, depending on whether the answer appears affirmatively, appears negatively, or is absent in the comprehension model.

Additionally to the above, the user may also select to present only part of the comprehension model, or have the system present the arguments that it used to support the inferences that led to the comprehension model. Roughly, the user may request to see: the ``universal argument'' showing all rules that are activated by the story (in all extensions) without regards to conflicts, the ``acceptable argument'' showing those activated rules that are accepted (in the grounded extension) after the argumentation semantics resolve all conflicts, and details on which rules are qualified (``attacked'') by other rules to help the user understand why certain rules did not end up in the acceptable argument. Since the comprehension model is revised from session to session, the user may also see which rules become obsolete and are retracted across sessions (i.e., part of the grounded extension for the preceding but not the current session), and which new rules come into play and are used to elaborate the comprehension model (i.e., part of the grounded extension for the current but not the preceding session). Finally, the user may choose to see how much time the STAR system spends in each part of its computation, and to decide whether the relevant part of the story will be shown along with each session.

More details about the STAR system can be found in the work of \citeN{Diakidoy2015} and on the system's website (http://cognition.ouc.ac.cy/star). The full example used in this section, is available from the ``Help'' menu of the Web-STAR IDE.

\section{Building Web-STAR on top of STAR}
\label{section:webstar}
Following the successful paradigm of many other projects that moved to an online environment, and aiming towards increasing the usage of the STAR system from non-expert users, we developed a web-based IDE for STAR. This IDE incorporates all the functionality of the STAR system in a structured web environment with the addition of visualization, automation, and collaboration tools that help users prepare and process their stories.

Moreover, the IDE employs a number of social features for user collaboration, like public code sharing and posting of stories to a public repository. In addition, users can work together using a state-of-the-art collaboration component which allows screen sharing, text and voice chat, and presenter following functionality. In short, work on Web-STAR includes:

\begin{itemize}
\item A Web-IDE that does not require setup, it is OS agnostic, and offers modern IDE functionalities.
\item A platform for collaboration and educational support.
\item A platform for integrating story comprehension functionality to other systems.
\item A modular architecture that facilitates the addition of new components and functionality.
\end{itemize}

\begin{figure}
\begin{center}
\includegraphics[width=1.0\linewidth]{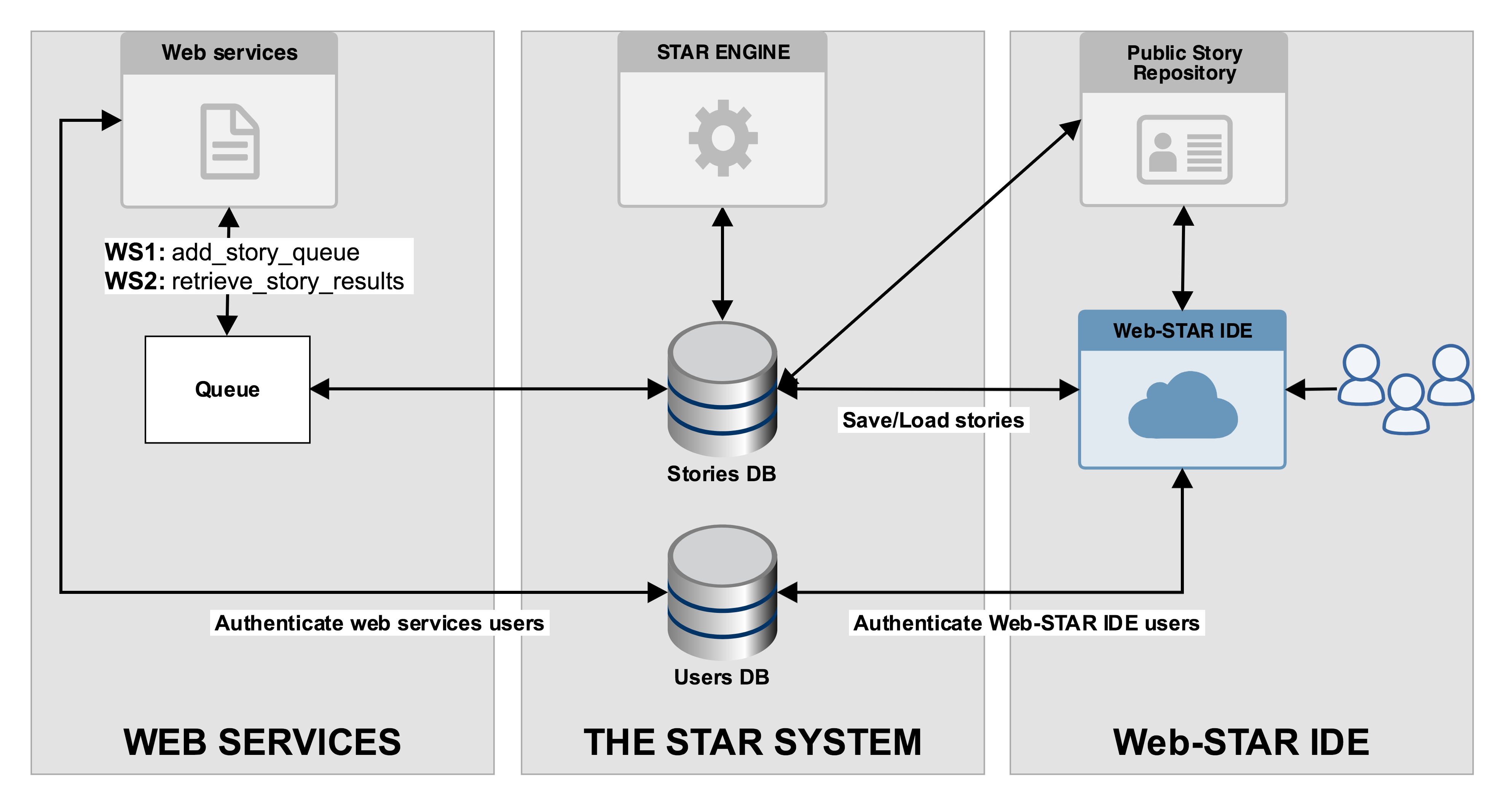}
\caption{The Web-STAR platform architecture with its three core components: the Web-STAR IDE, the STAR system, and the web services infrastructure. The diagram also presents the authentication mechanism, the storage functionality, and the web services provided.}
\label{fig:webstar:architecture}
\end{center}
\end{figure}

Fig.~\ref{fig:webstar:architecture} depicts the architectural diagram of the Web-STAR platform that comprises the Web-STAR IDE, the web services, the STAR system engine, the public repository, and the databases for storing related information. In the next paragraphs, a presentation of the web-based IDE is shown with details of the workspace layout, the components, and the functionality of the IDE. The Web-STAR IDE is available online at http://cognition.ouc.ac.cy/webstar/ and it is accessible from any device.

\subsection{Getting started with the Web-STAR interface}
\label{webstar:subsection:strting_webstar_interface}
To start using the Web-STAR IDE, a user creates an account and activates the personal workspace. Currently, both local and remote authentication options are available. The local authentication method uses the integrated storage facilities of the platform. The remote authentication method uses the OAuth2 protocol (https://oauth.net/2/), offered by third parties like Facebook, Google, Github, etc. Other authentication methods are supported as long as the appropriate plugin is available.

After the authentication process is completed, the user is redirected to the Web-STAR IDE environment where both the source code editors and the visual editors are present.

\subsection{The IDE environment and workspace}
\label{subsection:webstar:workspace}
\begin{figure}
\begin{center}
\includegraphics[width=1.0\textwidth]{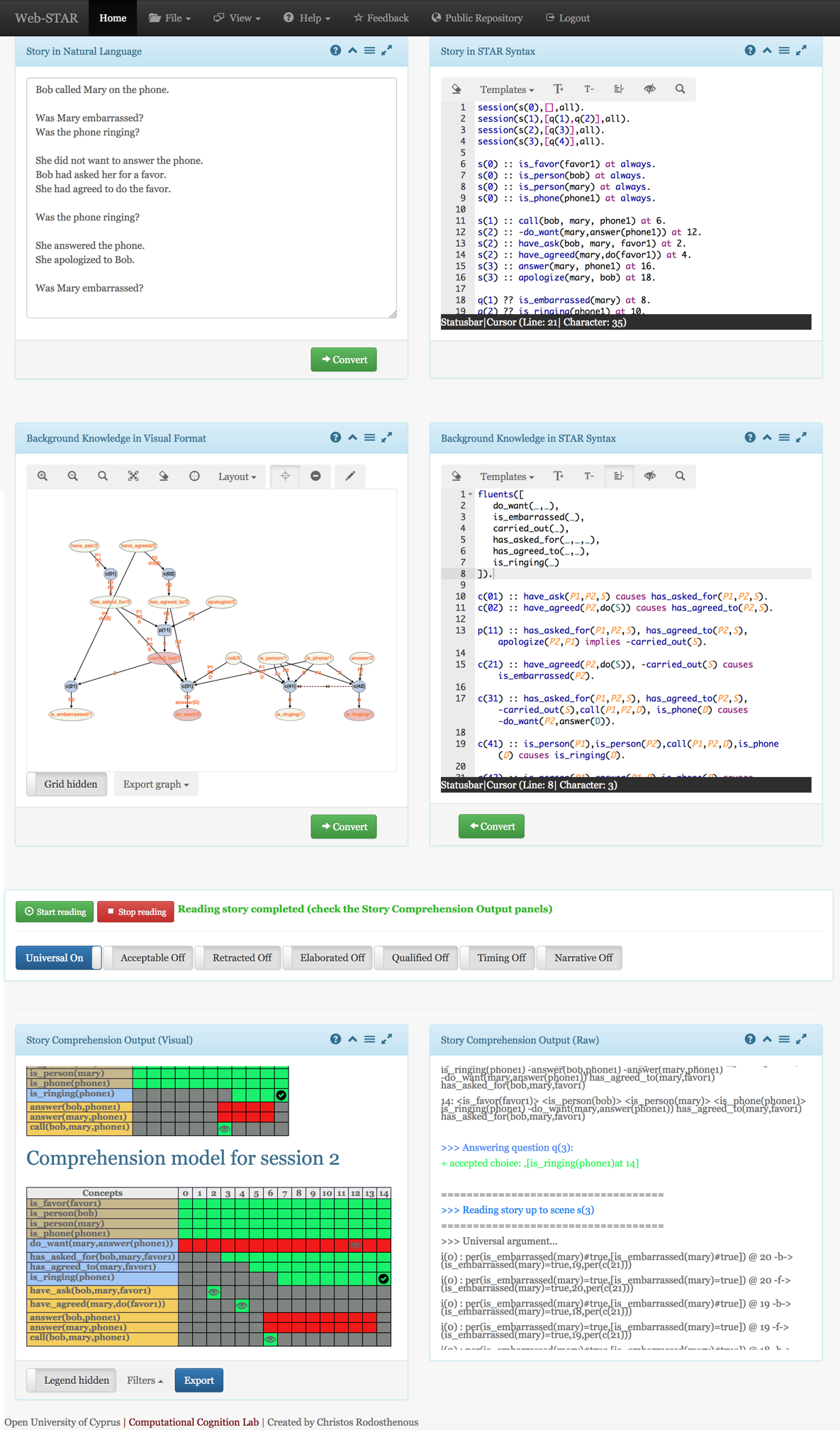}
\caption{A screenshot of the Web-STAR IDE layout. The workarea is divided into two columns: the left column (Simple mode) and the right column (Advanced mode); and three rows: the story area, the background knowledge area, and the story comprehension output area.}
\label{fig:webstar:view_modes}
\end{center}
\end{figure}

Users are presented with the workspace (see Fig.~\ref{fig:webstar:view_modes}), which is divided into three distinct areas: \begin{enumerate*}[label=(\roman*)] \item the story writing area, \item the background knowledge writing area, and \item the story comprehension output area \end{enumerate*}. This design was chosen to give users a clear understanding of the workflow of the story comprehension process, and to enable users to hide the areas which are not needed, aiming to avoid information overload.

The workspace is also divided into two columns. The left column is for the tasks that do not require users to have prior knowledge and experience in using the STAR system and the right column is for more seasoned users who have prior knowledge of the STAR system semantics and experience in encoding stories using the STAR system. This modular layout allows users to choose the mode they want to use while preparing their stories. More specifically, the web interface comprises three view modes:
\begin{itemize}
\item \textbf{Simple}: Users write a story in natural language, add background knowledge using the visual editor, and view the visual representation of the story comprehension model. This mode is ideal for users that are new to story understanding systems and want to have an overview of the capabilities and processes of encoding a story.
\item \textbf{Advanced}: Users write a story in the STAR syntax, encode the background knowledge in the source code editor, and are presented with raw output from the STAR reasoning engine. This mode is ideal for users with prior knowledge in encoding stories in the STAR syntax, and for users who wish to enter pre-encoded stories as done in the standalone version of the STAR system.
\item \textbf{Mixed}: Users write a story using any of the above options and convert from one mode to the other. For example, users can encode the background knowledge using the source code editor and then convert it to the visual format where they can make further changes. This mode is ideal for users that are learning the system and feel more confident in using the visual components and viewing the conversion. Moreover, this mode is used for teaching, since educators can present examples of encoding the story in visual format and then present the corresponding STAR syntax.
\end{itemize}

Additionally, users are also able to set the active area in any of the above view modes, i.e., to display only the background knowledge area, the story input area, or the story comprehension output area. The design of the workarea is fully customizable, allowing users to maximize the part of the IDE they are currently working on and minimize the areas that are not needed at that time. Whenever a mode is chosen, the relevant area is resized to maximize the view to the screen size of the device.

\subsection{The story workarea}
\label{subsection:story:workspace}
Users can start using the system by creating a story from scratch, either in natural language or in the STAR syntax, or even by loading an existing story. More specifically, users can write their code in the source code editor or load it from an external file previously created for the standalone STAR system, load an example file, or load a story file from the public repository. Non-experts can benefit from the example stories and the user-contributed stories in the public repository.

\begin{figure}
\begin{center}
\includegraphics[width=1.0\linewidth]{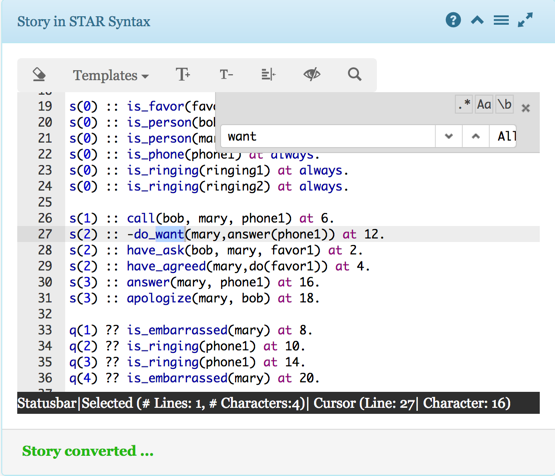}
\caption{A screenshot of the source code editor, depicting the line numbering, syntax highlighting and line highlighting functionality. Above the source code editor resides the toolbar menu with the ``search \& replace'' functionality window open. At the bottom of the editor resides the statusbar with information on the selected line and character.}
\label{fig:webstar:source_code_editor}
\end{center}
\end{figure}

Currently, the source code editor (see Fig.~\ref{fig:webstar:source_code_editor}) allows syntax highlighting using a STAR syntax highlighter file that inherits the Prolog's syntax template and is expanded with the STAR semantics. Furthermore, line numbering and code wrapping are also available to users along with the extensive ``search \& replace'' capability for finding text in large stories.

The Web-STAR IDE has a comprehensive list of menu options that enable users to load example story files, study them and edit them. Users have a personal workspace for saving their newly created stories and a public workspace for loading other users' stories. A story that is saved in the personal workspace can only be accessed by its creator, whereas a story stored in the public space is visible to everyone. Options for importing code stored locally on the user's personal device and exporting stories to a file for local processing are also available. This functionality allows a user to use the standalone version of the STAR system to process the story.

When a user loads a story, like the example story presented in Section \ref{section:star_system}, the source code editor immediately identifies and highlights the STAR semantics (variables, rules, operators) and makes it easier for the user to read the encoded story (see Fig.~\ref{fig:webstar:source_code_editor}). After studying the file, the user can move to the questions part and can add one or more questions by choosing the ``question template'' from the menu. The question template (presented in Section \ref{section:star_system}) is added and the user can add the predicates and time-points at which the question is posed. When changes are made to the example file, the user can save it to the personal workspace using the corresponding menu option.

\subsubsection{Natural language to the STAR syntax converter}
\label{subsection:nl_to_star}
One of the innovations available to the Web-STAR user is the automated component to convert a story from natural language to the STAR syntax. This is a real-time process, where the story and the questions are written in natural language, and the system processes them using a Natural Language Processing (NLP) system and a custom-built parser that maps processed words and phrases to predicates with their arguments.

More specifically, each sentence is processed using the Stanford CoreNLP \cite{manning-EtAl:2014:P14-5} system for NER (Named Entity Recognition), part-of-speech, lemmas (canonical or base form of the word), basic dependencies, and coreference resolution.

First, the component automatically identifies the sessions or scenes of the story. Sessions are added when a series of statements are followed by a question, or a group of questions. There is always a base session ``\textbf{Session S(0)}'' where all the constants are represented. For each group of questions, an additional session is created (e.g., see the story in section \ref{section:star_system}, Fig.~\ref{fig:example:story_star}).

Next, the nouns in each sentence and the named entity types (location, person, organization, money, percent, date, time) are identified to create the concepts that represent constant types. For each named entity, a statement of the form ``\texttt{is\_<EntityType>(<entity>) at always.}'' is added to the base session of the story. Personal pronouns are also identified as a \texttt{Person} entity. The following is an example of this: \texttt{is\_person(personX) at always.}, where X is an integer, representing the number of entities with the same name. When a coreference is found, the identified person name is used in the concept.

Predicates are created using the Stanford basic dependencies for extracting textual relations from the text. More specifically, for each sentence, the lemmatized ``\texttt{ROOT}'' is used as the predicate name and the lemmatized text from types ``nsubjpass, dobj, nmod:poss, xcomp'' is used to create the predicate arguments in lower case.

When a word is characterized with the ``aux'' or ``compound:prt'' types, then the predicate name is expanded with a ``\_'' and the new word is added in front of the predicate name. When a word has a ``neg'' type, then the negation symbol ``-'' is added in front of the predicate name. Words with types ``amod, case, cop, auxpass, aux and compound:prt'' that are dependent of the ``ROOT'' are also appended to the predicate name. For words with types ``aux, aux:pass and cop'' when the lemma ``be'' is identified, it is converted to ``is''.

For adding time-points to the story statements, we start at time-point 2 and form two lists. First a list with statements in past perfect is formed starting from time-point 2 and increasing by two for each statement. Then a list with statements starting from the maximum time-point of the first list with an increment of two is formed. These two lists are joined and form the story statements (see Fig.~\ref{fig:example:story_star}).

\begin{figure}
\begin{center}
\includegraphics[width=1.0\textwidth]{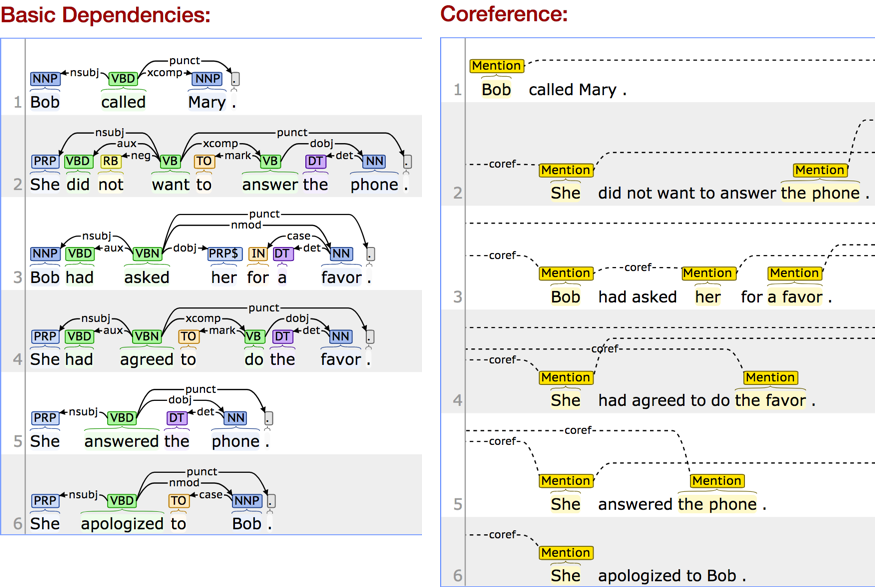}
\caption{The output of the CoreNLP processing for the example story in Fig.~\ref{fig:example:story_nl}. On the left side, the basic dependencies are presented in graphical form and on the right side the coreferences are depicted.}
\label{fig:core_nlp_example}
\end{center}
\end{figure}

To better understand the conversion process, we take the example of the story in Fig.~\ref{fig:example:story_nl}, its representation in the STAR syntax (see Fig.~\ref{fig:example:story_star}) and the output from the Stanford CoreNLP depicted in Fig.~\ref{fig:core_nlp_example}. In particular, for the sentence: ``She had agreed to do the favor'' (sentence 4 in Fig.~\ref{fig:core_nlp_example}) we take the ROOT (``agreed'') and the dependent words ``She'',``do'' and form the predicate \texttt{agree (She,do)}. Next, the ``ROOT'' is connected with an ``aux'' type with the word ``had'' and the predicate name is updated accordingly \texttt{have\_agreed(she,do)}. The word ``She'' refers to ``mary'' (using the coreference parsing) and ``do'' is connected with an ``xcomp'' relation with the \texttt{ROOT}, so ``do'' will form a new predicate \texttt{do(favor1)} and the final concept will become \texttt{have\_agreed(mary,do(favor1))}.

\subsection{The background knowledge workarea}
\label{subsection:story:bgknowledge}
\begin{figure}
\begin{center}
\includegraphics[width=1.0\textwidth]{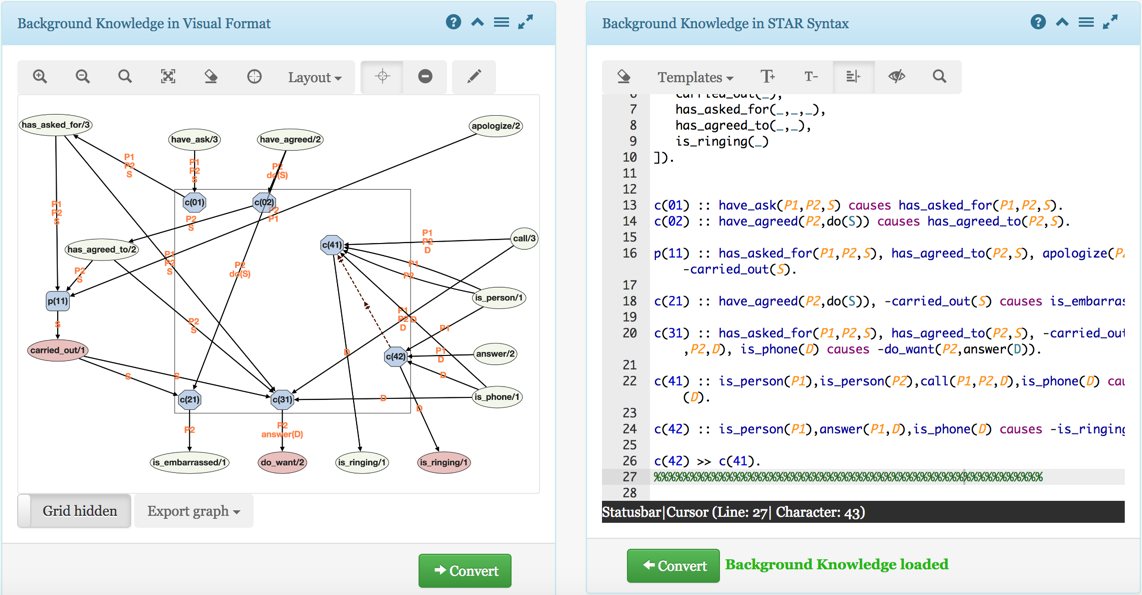}
\caption{The background knowledge workarea of the IDE. The visual editor is depicted on the left side of the screenshot, and the source code editor on the right side. There are conversion buttons from one form to the other at the bottom of each panel.}
\label{fig:bkeditor:workarea}
\end{center}
\end{figure}

The next step in preparing the story is the encoding of the background knowledge. This can be done either by using the source code editor or the visual editor (see the left side of Fig.~\ref{fig:bkeditor:workarea}). The visual editor uses a directed graph to represent rules. Users are able to see how rules build on each other (i.e., rules whose body literals are the head literals of other rules) and better understand the reasoning process. Moreover, users can focus on specific rules and literals and understand their role in forming the comprehension model of the story.

In particular, each rule is represented with a blue-colored node of octagonal shape for causal rules, and of a rounded orthogonal shape for property rules. Literals are represented with nodes of a cyclical shape, and are green or red to indicate that the literal is, respectively, positive or negative. Literals with a directed edge towards a rule node represent body literals for that rule, and the single literal with a directed edge from a rule node represents the head literal for that rule.

\begin{figure}
\begin{center}
\includegraphics[width=0.6\textwidth]{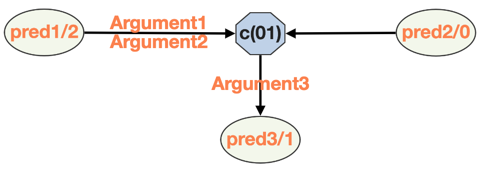}
\caption{A visual representation of the rule: \texttt{c(01) :: pred1(Argument1,Argument2), pred2 causes pred3(Argument3).}}
\label{fig:bkeditor:graph_rule_repr}
\end{center}
\end{figure}

Each node is labeled with the rule's or the predicate's name. For literals, the name is created using the predicate's name and the arity of the predicate (e.g., the predicate ``have ask'' with three arguments is represented as \texttt{have\_ask/3}). Arguments are represented with labels on the edges connecting the literal nodes with the rule node (see orange labels on edges in Fig.~\ref{fig:bkeditor:graph_rule_repr}).

Users can choose to create the entire background knowledge using the tools of the visual editor. Using the ``edit'' button, users can add literals, rules, edges, and priorities between rules. More specifically, users choose the desired element and click on the white area of the graph, the ``canvas''. When a rule is added, the label is automatically set to create a unique name (e.g., \texttt{c01, c02, p01, p02 \dots}). When literals are added, the user is asked to provide the literal's name, arity, and polarity (positive or negative). Users connect literals with rules using the ``edge drawing tool'', and can set priorities between rules by drawing a ``dashed edge'' from one rule node to another. Adding and updating arguments to literals is performed by clicking on the connecting edges between the literal and the rule. A dialog box appears for typing each argument.

For every input (textual or visual) and every action on the canvas that does not conform with the STAR syntax (e.g., having two literals in the head of a rule) a guidance message (not simply an error message) is shown, which explains to the user in a visual way (e.g., by highlighting nodes or edges) what needs to be changed to lift the error. This is helpful in teaching scenarios, where students get to know the environment and the basic semantics of the STAR system (or logic-based programming, more generally).

\begin{figure}
\begin{center}
\includegraphics[width=1.0\textwidth]{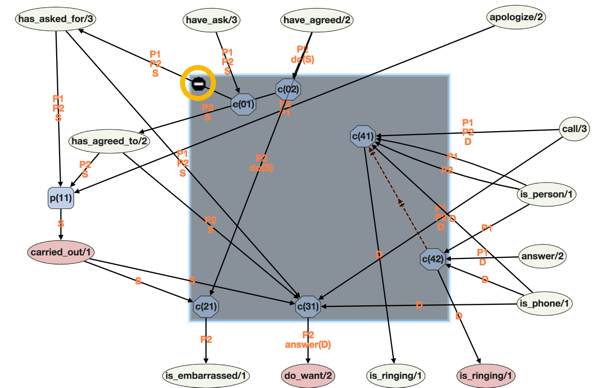}
\caption{A screenshot of the folding/unfolding code capability of the Web-STAR IDE. Grouped rules can be maximized and minimized by clicking at the top left corner of the grouping.}
\label{fig:bkeditor:folding_unfolding}
\end{center}
\end{figure}

In cases of stories with a large background knowledge, a user can group rules together and minimize or maximize the view of individual groups, isolating the part of the background knowledge that the user wants to inspect. To support this, a type of code folding functionality is implemented, which allows users to focus on a specific subset of the rules on the screen. In Fig.~\ref{fig:bkeditor:folding_unfolding} the code folding/unfolding capability of the IDE is presented.

Moreover, users can zoom in and out of the graph and can change its layout dynamically. There are a number of available layouts for users to choose from, like ``the circle layout'' where nodes are put in a circle, ``the breadthfirst layout'' where nodes are put in a hierarchy based on a breadthfirst traversal of the graph, etc. Furthermore, users can search for a rule, literal or argument using the search tool. When the element is found, it is maximized and focused along with its neighboring elements. There is also an option for fitting the graph to the screen, as well as a ``graph navigator'' option that allows users to have a bird's eye view of the whole graph and navigate to the desired part of it.

\begin{figure}
\begin{center}
\includegraphics[width=1.0\textwidth]{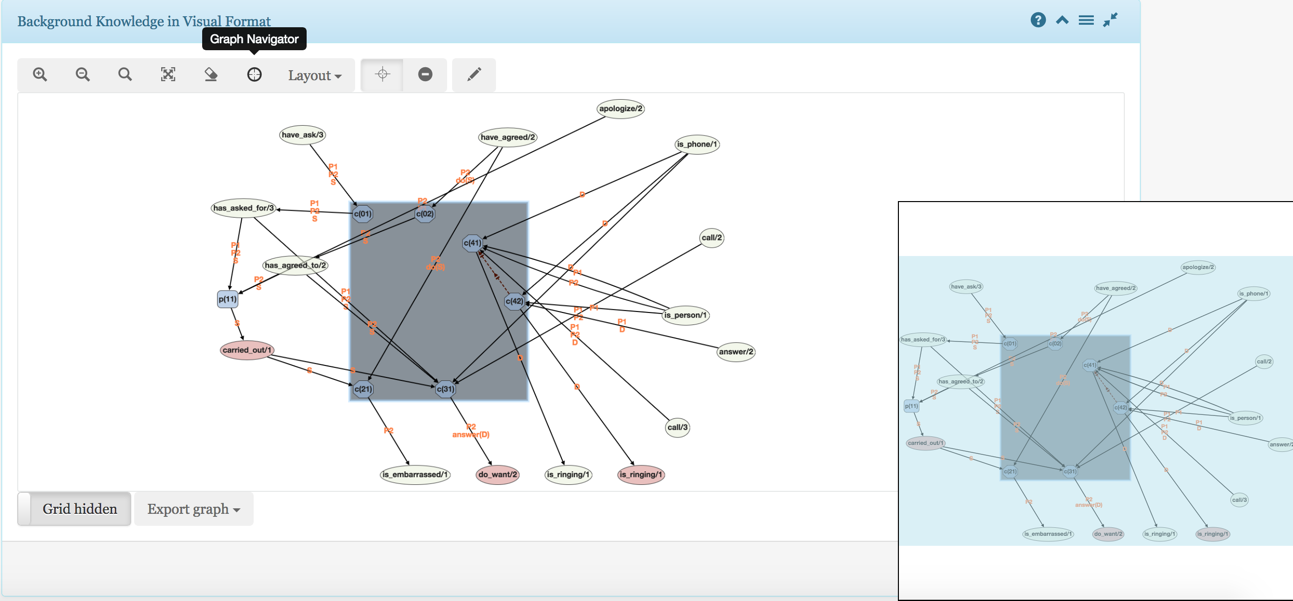}
\caption{A screenshot of the ``graph navigator'' window (bottom right), where users can have a bird's eye view of the background knowledge graph and navigate to the desired part of it.}
\label{fig:bkeditor:graph_navigator}
\end{center}
\end{figure}

The Web-STAR IDE allows filtering out elements of the graph that are not needed for a specific job. For example, users can choose to ``toggle'' the visibility of causal or property rules, priorities, rules with low density or rules that are not connected with other rules. This functionality is part of the Web-STAR's IDE ability to handle large background knowledge bases with rules.

The background knowledge graph can be exported in various formats, including image formats (png, jpg), JSON, and GraphML \cite{Brandes2013Graph-25913}, which can subsequently be used with third-party applications to present or process the graph and its data.

Expert users can use the source code editor in parallel with the visual one. A similar in look and feel editor with the one for preparing stories is available, and users can take advantage of the included templates for adding rules.

\subsubsection{Converting background knowledge from visual to textual format}
\label{subsection:convert_bg_knowledge}
Converting background knowledge from visual to textual format and vice versa, is performed with the click of a button, allowing the user to encode parts of the knowledge in the one format or in the other. Web-STAR's internal mechanisms read each graph element and perform the conversion of visual background knowledge rules to STAR format. Details on this process are provided in Algorithm \ref{algorithm:convert:graph_to_text}.

\begin{algorithm}[h]
\caption{Convert background knowledge graph to the STAR syntax}
\label{algorithm:convert:graph_to_text}
\begin{algorithmic}[1]
\Procedure{convert\_graph\_to\_STAR(GRAPH\_OBJECT)}{}
\State \% Get all the rule nodes of the graph object

\For{Each node $i$}
\State \%Get the edge directed outwards of the node (Head)
\For{Each edge $j_1$}
\State literal=edge[$j_1$].connected\_node
\State head=convert\_predicate\_star(literal,edge[$j_1$].label)
\State push\_to\_list\_of\_head\_literals(head)
\EndFor
\State \%Get all the edges directed towards the node (Body)
\For{Each edge $j_2$}
\State literal=edge[$j_2$].connected\_node
\State body=convert\_predicate\_star(literal,edge[$j_2$].label)
\State push\_to\_list\_of\_body\_literals(body)
\EndFor
\State \%Proceed and create the textual representation of the rule
\If{node[$i$].type===``property''}
\State rule\_type=``implies''
\Else
\State rule\_type=``causes''
\EndIf
\State rule=node[$i$].name :: node[$i$].body\_literals rule\_ type node[$i$].head\_literal
\EndFor
\EndProcedure
\end{algorithmic}
\end{algorithm}

\subsection{Story comprehension process and output}
\label{subsection:story_comprehension}
After completing the story preparation and the background knowledge encoding, users can proceed with the story comprehension process. Users can click the ``Start reading'' button and immediately see results coming from the STAR system in the ``Story Comprehension Output'' area in real-time. A number of reporting options can activate and expose the internal processes of the STAR system, including the argumentation mechanism applied for story comprehension, for debugging or educational purposes. In particular, users can choose to view all arguments (\texttt{Universal}), the subset of acceptable arguments (\texttt{Acceptable}), arguments removed during a specific session (\texttt{Retracted}), arguments added during a specific session (\texttt{Elaborated}), and information about which arguments qualify other arguments (\texttt{Qualified}).

When the reading process is completed, users can view both the comprehension model and the answers to questions posed. This can be done both in a visual and textual format. The visual output might be preferred for tracking each concept across the story time-line, and the textual output might be preferred for debugging. Each panel is dynamically updated when new information is sent from the STAR system.

\begin{figure}
\begin{center}
\includegraphics[width=1.0\textwidth]{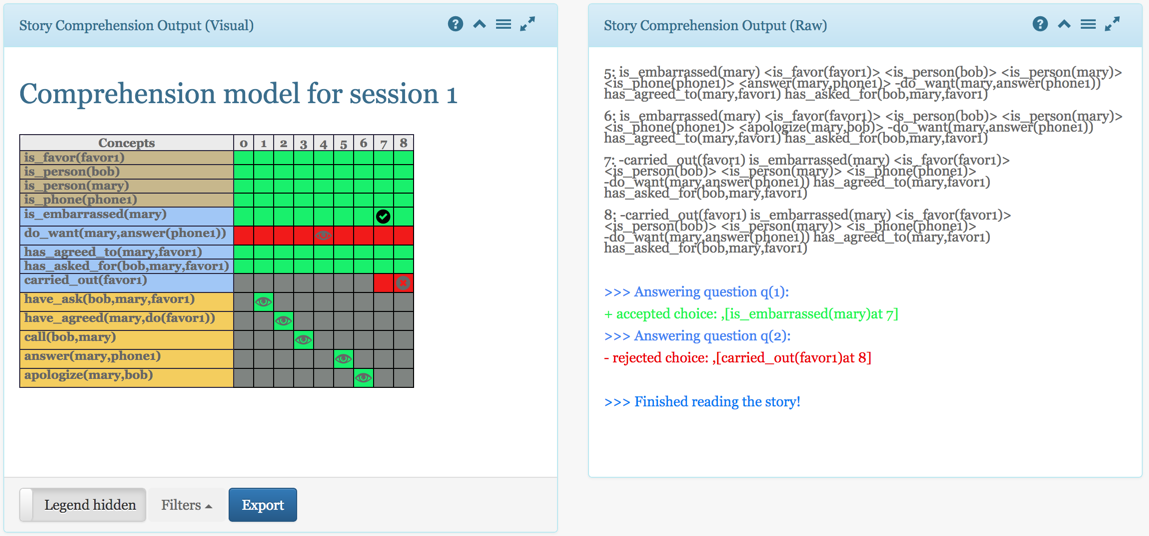}
\caption{A screenshot of the ``Story comprehension output'' workarea of the IDE, depicting the comprehension model. The legend above the comprehension model provides details on the meaning of symbols and colors in the visual representation of the model and its visibility can be toggled by using the relevant switch. On the right side, the raw output for the same story is presented.}
\label{fig:webstar:live_reasoning}
\end{center}
\end{figure}

In Fig.~\ref{fig:webstar:live_reasoning}, the visual output of the comprehension model is depicted, presenting the state of each concept at each time-point. Green, red, and dark grey represent concepts whose value is, respectively, positive, negative, or unknown at that time-point

The magnifying glass marks concepts whose value is observed at that time-point, i.e., they are extracted from the narrative directly. Concepts with orange background, represent an instantaneous action. Concepts with light blue background represent a persisting fluent and concepts with purple background represent a constant type (e.g., \texttt{person(bob) at always.}).

Users can apply filters on the output of the comprehension model and focus their attention of particular concepts. They can choose, for instance, to filter out fluents, actions, and constants, or to view only concepts whose value changes through time, concepts that have a high frequency in the background knowledge, or even concepts that are part of causal rules. The latter are a good indication of the focus of the story and its parts that are most interesting to a reader \cite{goldman1999narrative}.

The model can be exported in various formats and can be used for educational purposes. The Web-STAR IDE has also a textual format of the story comprehension output that presents the raw output of the STAR system as it would appear when executed as a standalone application. This output is enhanced with color highlighting to identify questions, positive or negative answers to questions, and debugging messages.

\subsubsection{Collaboration and ``social'' options}
\label{subsection:collaboaration}
\begin{figure}
\begin{center}
\includegraphics[width=1.0\textwidth]{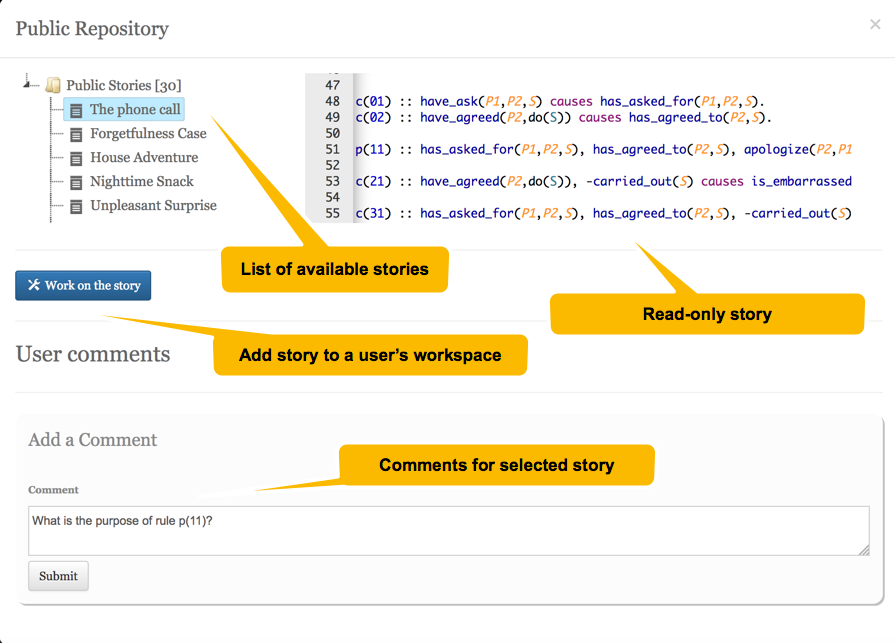}
\caption{A screenshot of the Public Stories Repository. Users can add comments on stories and ask questions. If interested, they can start working on a story by copying it to their personal workspace.}
\label{fig:webstar:public_repository}
\end{center}
\end{figure}

Apart from the typical IDE functionality, Web-STAR IDE also provides functionality for sharing publicly a story with other users. By clicking the ``Share it'' button a story is added to the public stories repository (see Fig.~\ref{fig:webstar:public_repository}) and appears in the ``public stories'' tab in the story browser dialog. Users can read shared stories and add comments, supporting the education of new users from more expert ones.

Beyond sharing, users can collaboratively write a story using the collaboration functionality provided. The system produces a link that can be sent to anyone interested in collaborating for a specific session. The recipient of the link can see the screen and the mouse pointer of each participating user, and changes of content in real-time, while also being able to chat through text and audio. This setting enables teams to collaborate on preparing a story, and allows students to learn by working together on class projects.

\subsubsection{User support and feedback}
\label{subsection:help_feedback}
The Web-STAR IDE offers a number of features to help its users achieve their goals. Firstly, users can follow a guided tour through the Web-STAR IDE features. Users can then start testing the functionality of the platform with the examples available in the story browser. These examples were carefully crafted for teaching the STAR semantics.

Moreover, in each panel there is an online help option, for guiding users to the specific functionality available for that panel. The icons and graphics chosen for buttons and toolbars are inline with the ones users are familiar in other IDEs. In cases where some users are not aware of the meaning of an icon, a tooltip is available.

To allow users to provide feedback on new desired functionalities or encountered problems, Web-STAR offers a built-in feedback functionality that stores a user's message in the platform's database and alerts the developers through email.

\subsection{Technical details and challenges}
\label{subsection:technical_details_and_challenges}
For designing and implementing the Web-STAR IDE, we chose to use technologies that are mature, do not require license fees, have a large community of contributors, and can be deployed easily. Furthermore, all technologies used are available as free and opensource software, and their communities release frequent updates and new capabilities in each new release. These considerations are important for a project that seeks to be expandable, scalable, and easy to maintain.

The system is based on PHP for backend operations, on the MariaDB database for the data storage, and on the JQuery JavaScript library for the front-end design. Behind this infrastructure lies the STAR system \cite{Diakidoy2015} and the SWI-Prolog \cite{wielemaker_schrijvers_triska_lager_2012} interpreter. A wrapper is employed for sending the story file from the front-end to the back-end and returning the results in real-time from the Prolog interpreter using the HTML5 ``Server-Sent Events'' functionality to dynamically update the interface.

All data storage is handled with the MariaDB database. In particular, a number of tables are used for storing user data, user profiles, and the STAR web service queue.

For the interface design, the Bootstrap framework is used. Bootstrap is an HTML, CSS, and JS framework for developing responsive projects on the web. This framework has a number of ready-to-use components like buttons, panels, toolbars, etc., and is also supported by a large community that develops extra components. The JQuery library (https://jquery.com/) is used to add intuitive UI components and AJAX functionality.

Collaboration functionality is provided using both AJAX components for sharing and commenting on stories, and the TogetherJS library (https://togetherjs.com/). TogetherJS is a JavaScript library from Mozilla that uses the Web RTC \cite{Johnston:2012:WAR:2432294} technology to enhance communication. It provides audio and chat capabilities between users, and allows users to see each other's mouse cursors and clicks, and the screen content.

The source code editor is based on the ACE editor (https://ace.c9.io), an open source web editor which is used by many other popular cloud IDEs. This editor was chosen because of its maturity, its open source license, and for its popularity. ACE is a code editor written in JavaScript and includes features like syntax highlighting, theming, automatic indent and outdent of code, search and replace with regular expressions, tab editing, drag-drop functionality, line wrapping, and code folding. Moreover, this editor can handle huge documents with more than one million lines of code.

For the visualisation of the background knowledge, we sought a component that is able to represent rules in a graph format and can additionally allow interaction with the user and the graph elements. For that reason, Cytoscape.js \cite{cytoscape2016} was selected, which is an open-source JavaScript-based graph library that allows users to interact with the graph, supports both desktop browsers and mobile browsers. It can also handle user events on graph elements like clicking, tapping, dragging, etc. This library also provides a large number of extensions that are employed to enhance the functionality of the Web-STAR IDE. The code folding/unfolding capability uses the ``expand-collapse'' extension, which provides an API for expanding and collapsing compound parent nodes on a cytoscape graph. The ``edge drawing'' tool of the visual editor uses the ``edgehandles'' extension, which provides a user interface for dynamically connecting nodes with edges. The graph navigator capability is based on the ``navigator'' extension, which provides a bird's eye view with pan and zoom control from the graph.

For converting a story from natural language to the STAR syntax, we use a custom-built component developed at our lab, which uses the Stanford CoreNLP for natural language processing, a python script for processing the NLP output, and PHP for orchestrating and delivering the results through a RESTful API. The Web-STAR IDE integrates this component into its workflow, while the same methodology can be used by other systems to acquire this functionality.

The Web-STAR platform exposes two web services that can be used by third party applications, for adding a domain file to the STAR system queue in order to process, and for retrieving the results after the completion of the reasoning process (see Fig.~\ref{fig:webstar:architecture}): the ``\texttt{add\_story\_queue}'' web service takes as a parameter the story in the STAR syntax and returns a unique identifier; the ``\texttt{retrieve\_story\_results}'' web service takes as a parameter the unique identifier previously sent by the ``\texttt{add\_story\_queue}'' web service, and returns the results of the comprehension process. This approach was chosen to minimize the waiting time in cases of large story files that require extensive processing.

\section{Web-STAR IDE evaluation}
\label{section:evaluation}
An important step in the design and deployment of a web-based IDE is the evaluation of its usability, i.e., ``the degree to which users are able to use the system with the skills, knowledge, stereotypes, and experience they can bring to bear'' \cite{eason2005information}. Usability evaluation can be conducted using interviews, task analysis, direct observation, questionnaires, and heuristic evaluation, among others \cite{Barnum:2001:UTR:559526}. In terms of evaluating an IDE, \citeN{Kline2005} presented three techniques which can also be applied for the Web-STAR IDE's evaluation: \begin{enumerate*}[label=(\roman*)] \item the unstructured interviews, \item the heuristic evaluation and psychometric assessment, and \item the laboratory observation combined with the cognitive walkthrough \end{enumerate*}. Moreover, in work by \citeN{pansanato2015evaluation}, the capturing of user interaction is stretched for usability evaluation of rich web interfaces. The authors present a number of tools and methods that go beyond simple capturing of log files from the web server, like the recording of user interaction from the client side, i.e., the browser.

\subsection{Evaluation setting}
We followed a hybrid approach for the Web-STAR IDE's evaluation that combines the cognitive walkthrough method \cite{John:1995:LUC:223904.223962,Blackmon:2002:CWW:503376.503459} with questionnaires and user interaction capturing techniques. The process was divided into the design phase, the pilot phase, and the actual evaluation phase, and sought to:
\begin{itemize}
\item Evaluate the web-interface in terms of ease of use, understanding, learnability, and efficiency.
\item Detect possible usability problems of the Web-STAR IDE.
\item Perform the above for both experts and non-experts that use the IDE.
\end{itemize}

\subsubsection{Design phase}
\label{subsection:design_phase}
The design phase involved the selection of the participants for the evaluation, the design of the tasks that each participant would undertake, the preparation of the questionnaires, and the technical methods for tracking each participant's interaction with the system.

Participants were chosen from both groups that would have an interest in using the Web-STAR IDE: \begin{enumerate*}[label=(\roman*)] \item experts, and \item non-experts \end{enumerate*}. The expert group included computer scientists and psychologists with prior experience in using the STAR system as a standalone Prolog application, and computer scientists or computer science students with programming skills in Prolog or other declarative programming languages. The non-expert group included psychologists, school teachers, law students, and students of psychology, who had very little or no experience in using IDEs or programming languages. A total of 15 participants were selected, which, according to the relevant bibliography \cite{macefield2009specify}, is an appropriate sample for detecting the majority of the usability problems of a system.

We compiled a list of \textbf{Cognitive Walkthrough Tasks} that were specifically designed to evaluate the major functions and aspects of the Web-STAR IDE. Each task instructed the user to perform a sequence of actions, as follows:
\begin{itemize}

\item \textbf{Task 1 (Create an account)}: Navigate to the Web-STAR IDE link and create an account. Activate the account and log into the system.

\item \textbf{Task 2 (Follow the guided tour)}: Follow the guided tour to learn the basic functionality of the IDE.

\item \textbf{Task 3a (Write a new story in natural language)}: Write a given story along with its questions in natural language and convert it to the STAR syntax. Then add the background knowledge given in a visual format, using the visual editor and convert it to the STAR syntax. Save the story.

\item \textbf{Task 3b (Write a new story in the STAR syntax)}: Write a given story along with its questions in symbolic format, using the source code editor. Then add the background knowledge given in the STAR syntax, and convert it to the visual format. Save the story.

\item \textbf{Task 4 (Load a story and initiate the comprehension process)}: Choose a public story, load it, and initiate the story comprehension process.

\item \textbf{Task 5a (Modify the background knowledge using the visual format editor)}: Load a story, add a new background knowledge rule given in visual format, update and remove an existing rule, all using the visual editor. Finally, initiate the comprehension process. 

\item \textbf{Task 5b (Modify the background knowledge using the source code editor)}: Load a story, add a new background knowledge rule given in the STAR syntax, update and remove an existing rule, all using the source code editor. Finally, initiate the story comprehension process.

\item \textbf{Task 6 (Filter the output of the comprehension process)}: Load a story, initiate the comprehension process, and filter the output to present only the concepts that change while the story unfolds.

\item \textbf{Task 7 (Share a story)}: Load a story and share it in the public story repository.

\item \textbf{Task 8 (Comment on a user's story)}: Find a story in the public story repository and add a comment on that story.

\item \textbf{Task 9 (Initiate the collaboration tool)}: Initiate the collaboration functionality, and send the generated collaboration link to another person using the feedback option.
\end{itemize}

Since not all participants are experts in encoding stories in a symbolic language, one of the major considerations when preparing these tasks was to obtain comparable results from the users. Towards that aim, both the story and the background knowledge in Tasks 3a and 3b were provided in the instructions given to the users.

After observing each participant performing the above tasks, the experimenters tried to answer the following \textbf{Cognitive Walkthrough Questions}, as explained in the work of \citeN{Wharton:1994:CWM:189200.189214}:
\begin{itemize}
\item Does the user try to achieve the right effect?
\item Does the user notice that the correct action is available?
\item Does the user associate the correct action with the effect that the user is trying to achieve?
\item If the correct action is performed, does the user see that progress is being made toward the solution of the task?
\end{itemize}
When the answer to any of these questions was ``No'', an error was counted towards the total number of errors for that task.

The time needed to complete each task was calculated from the time the participant logged into the IDE until the time the participant logged out of it, with an exception in the first task where the time was measured from the time the participant clicked the register button until the time the participant logged out of the IDE.

After the completion of the tasks, a \textbf{Demographics Questionnaire} was completed by participants to record their gender, age, degree, occupation, previous experience in using IDEs, knowledge of programming languages, and prior experience in using story understanding systems and more specifically the STAR system. A \textbf{Post-task Questionnaire} was also completed to capture the participants' opinion for using the IDE for each specific task. The questionnaire included questions that covered the various parts of the system invoked for each task: the interface (e.g., menu bar, panels, dialogs, buttons, and labels), the online help material, the visual editor, the public story repository, and the outcome of the story comprehension process. It included true/false questions, multiple choice questions, and questions in the five-point Likert scale. The questionnaire is shown in Appendix B. The questionnaire also included a section with questions from the \textbf{System Usability Scale (SUS) standardized questionnaire} \cite{brooke1996sus}, a ten-item questionnaire using a five-point scale for the assessment of perceived usability \cite{doi:10.1080/10447318.2015.1064654}. Both surveys were designed and deployed online and access to them was restricted to participants of our evaluation.

Finally, we implemented a logging functionality to capture detailed information from the participants' interaction with the Web-STAR IDE (e.g., login, menu selection, button click, visual editor usage) and measure the time between these interactions. This functionality was seamlessly integrated with the Web-STAR IDE using AJAX technology. Each event was stored in a database table and included the user-id of the participant that performed the action, the time the action was performed, the component used (e.g., login screen, menu bar, visual editor), the action (e.g., button click, visual editor graph node added), the data sent, and the response of the IDE.

The metrics chosen for the evaluation were both qualitative (e.g., user satisfaction, ease of use) and quantitative (e.g., number of successfully performed tasks, task completion time, number of errors occurred, number of times participants used the online help functionality, number of times participants clicked on a control).

\subsubsection{Pilot phase}
\label{subsection:pilot_evaluation}
Before the actual evaluation phase, we performed a pilot evaluation identical to the actual one, but with only two users, to verify that all tasks are feasible and understandable. This also allowed us to test that data were recorded properly and to get familiar with the testing process.

The pilot evaluation was performed in a laboratory environment with a computer connected to the Internet with access to the Web-STAR IDE URL. We also set up the screen recorder software to capture all interactions of the user with the interface (e.g., keystrokes, mouse movements, information dialogs, and visual editors) in a video file.

Both participants ended up needing more than an hour to complete the tasks and respond to the questionnaires.

\subsubsection{Evaluation phase}
\label{subsection:evaluation_phase}
All participants in the evaluation phase performed the experiment in a controlled environment which included a laptop with an Intel CORE i7 processor, 4GB of RAM, and a 15.4 inches screen. An external mouse was attached to the laptop, and participants had instructions to use it (instead of the laptop's integrated mousepad). The laptop was constantly plugged into the power source. In terms of software, the Firefox web-browser was used to load the WebSTAR IDE interface, and the Camtasia screen recording software was activated before each session to record the participant's actions. Each session was performed in a quiet room with only the participant and the experimenter present, aiming to minimize outside interference and noise from the environment.

As a first step, each participant completed a statement of informed consent regarding the reason for the evaluation, and the data collection and data handling policy. This consent was mandatory for participating in the evaluation. The participants were then asked to complete the Demographics Questionnaire online.

Following that, participants were presented with a document listing the Cognitive Walkthrough Tasks (see Appendix A). During the cognitive walkthrough, participants had continuous access, through the Web-STAR IDE, to online help files provided by the IDE, the STAR syntax guide, and the guided tour; i.e., the same type of help that any typical user of the IDE would have available while using it. Participants had the option to choose any type of viewing mode they saw fit when completing the tasks.

During each task, the experimenter recorded all observations made in a notebook, answered the cognitive walkthrough questions, and recorded problems and errors occurred while the participant used the IDE. After each task, the participant was presented with the post-task questionnaire for that specific task. The experimenter avoided providing any kind of verbal or non-verbal additional help to the participant while conducting the cognitive walkthrough.

After the completion of all the tasks, participants were presented with the System Usability Scale (SUS) standardized questionnaire. Finally, the experimenter stopped the screen recording, stored the capture, and saved all questionnaire answers online for later processing.

\subsection{Evaluation results}
\label{subsection:evaluation_results}
\begin{figure}
\begin{center}
\includegraphics[width=1.0\textwidth]{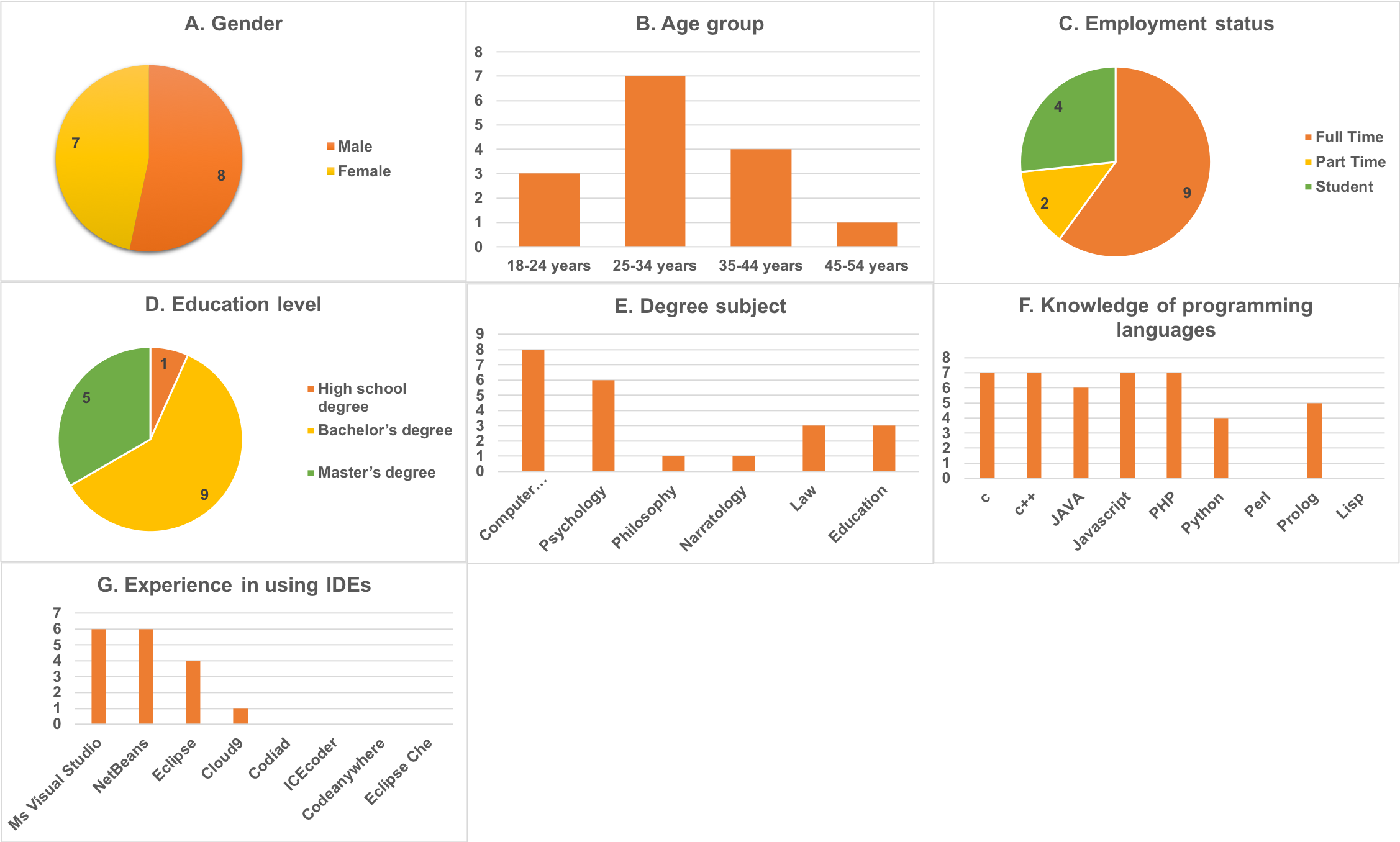}
\caption{Participants' demographics. Graph A depicts their gender, Graph B their age group distribution, Graph C their employment status, Graph D their education level, Graph E their degree subject, Graph F their knowledge of programming languages, and Graph G their experience in using IDEs.}
\label{fig:evaluation_analysis:demographics}
\end{center}
\end{figure}

Fifteen people (8 male, 7 female) participated voluntarily to the evaluation. All had Greek as their mother tongue and reported to have a normal or corrected-to-normal vision. All participants completed the whole evaluation process. Fig.~\ref{fig:evaluation_analysis:demographics} represents analytics regarding their gender, age group, education, employment status, and knowledge of programming languages and IDEs. More than half of the participants (8 out of 15) reported that they had heard the notion of story understanding, but only 2 reported that they had used a story understanding system before. In both cases, this system was the STAR system. The group of non-experts included 10 participants and the group of experts 5.

\begin{figure}
\begin{center}
\includegraphics[width=1.0\textwidth]{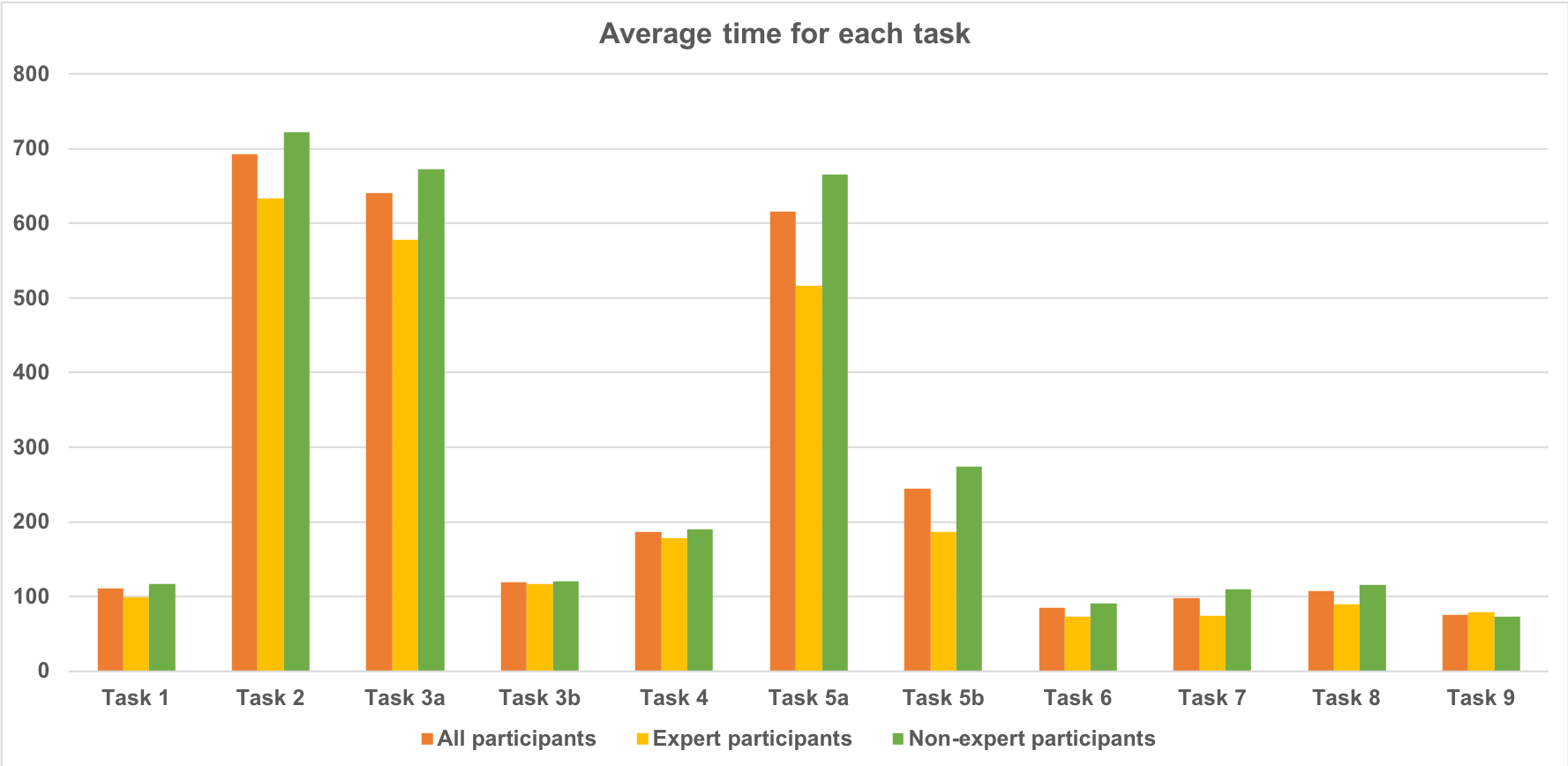}
\caption{Average time in seconds per task, for experts, non-experts, and all participants.}
\label{fig:evaluation_analysis:task_time}
\end{center}
\end{figure}

After the completion of the evaluation, the notes taken by the examiner for each participant with answers to the Congitive walktrough questions were carefully examined along with the answers in the post-evaluation questionnaire. The log files of each participant were also analyzed, and the aggregated results are presented in Table \ref{table:cognitive_walkthrough} and Fig.~\ref{fig:evaluation_analysis:task_time}.

\begin{table}
  \caption{Performance at the Cognitive Walkthrough evaluation}
  \label{table:cognitive_walkthrough}
  \begin{minipage}{\textwidth}
    \begin{tabular}{lcccccc}
      \hline\hline
      Task& Completed\footnote{Percentage of participants that successfully completed the task.}&
      Avg\footnote{Average time (in seconds) needed to complete the task.}&
      Std\footnote{Standard time deviation (in seconds) needed to complete the task.}& Max\footnote{Maximum time (in seconds) needed to complete the task.}&
      Min\footnote{Minimum time (in seconds) needed to complete the task.}&
      Errors\footnote{Number of errors recorded by the experimenter during the task.}\\
      \hline
      Task 1& 100\%&   111&  45&   202&   37&   0\\
      Task 2& 100\%&  692& 269&   1303&  224&   0\\
      Task 3a& 100\%&   640&  168&   1016&  387&   0\\
      Task 3b& 100\%&   119&  23&   156&  74&   0\\
      Task 4& 100\%&   186&  95&   381&   100&   0\\
      Task 5a& 100\%&  615& 105&   861&  438&   0\\
      Task 5b& 100\%&  244& 58&   338&  164&   0\\
      Task 6& 100\%&   85&  19&   114&  54&   0\\
      Task 7& 100\%&   98&  40&   186&   41&   0\\
      Task 8& 100\%&  107& 42&   230&  56&   0\\
      Task 9& 100\%&  75& 23&   132&  38&   0\\
      \hline\hline
    \end{tabular}
    \vspace{-2\baselineskip}
  \end{minipage}
\end{table}

The average total time for completing the evaluation tasks from both groups satisfy the normality assumption based on the Kolmogorov-Smirnov test. On average, experts performed the tasks of the cognitive walkthrough in less time (M=2622.40 seconds, SE=107.44) than non-experts (M=3150.00 seconds, SE=98.01), and the difference was significant \textit{t}(13)=3.32, \textit{p}$<$.05.

Further analysis of the results per task gives more insights on how participants interacted with the system (see Table \ref{table:cognitive_walkthrough}).

\subsubsection{Results from the cognitive walkthrough process}
In the following paragraphs we present the findings of the cognitive walkthrough process:

\textbf{Task 1}: All participants completed this task successfully. Some participants did not receive the activation email immediately due to email provider delays. The majority of participants clicked the links included in the introductory text while some others chose the ``Register'' option from the menu bar. It was also common for participants to try and press the enter button after filling up their credentials, but it was not working and they needed to click the ``Login'' button instead to proceed.

\textbf{Task 2}: All participants completed this task successfully. The majority of participants initiated the guided tour using the assistant dialog. A small number of participants did not understand that the assistant's message is clickable, and used the ``Help menu'' to find and start the guided tour. Some participants interacted with the IDE while following the guided tour and performed actions like ``reading a story'', ``drawing background knowledge'' using the visual editor and converting both the story and the background knowledge to test it. There were cases of participants who went back to a specific step, to test a functionality mentioned later in the guided tour. Moreover, one participant also expressed the opinion that it would be very useful if there was an option to watch a video instead of the guided tour. When the guided tour was showcasing the output panel, participants expected to have the story comprehension output area filled up with story information, but it was empty, since the story comprehension process was not activated by the guided tour.

\textbf{Task 3a}: All participants completed this task successfully. Some participants watched the help video first to properly perform the task. Participants used the visualizations, e.g., highlight of literals and rules which can be connected with an edge while drawing it, and red highlighting for incomplete rules along with debugging messages. Some participants were double clicking on the nodes and edges to move them and add arguments, when only a single click was needed.

\textbf{Task 3b}: All participants completed this task successfully. Participants found easily where to add the story and the background knowledge. A number of participants who tested the toolbar of the source code editor used functions and controls like the ``text wrap'' and the ``font size''.

\textbf{Task 4}: All participants completed this task successfully. Some participants had difficulties finding the ``Read Story'' button and they tried to locate it on the menu bar or on the top area of the IDE. Furthermore, non-experts read the confirmation message to understand where they could find the story output and when the reading process was completed.

\textbf{Task 5a}: All participants completed this task successfully. Some non-experts deleted the rule, but they forgot to delete the connected literal, whereas experts deleted the connected literal along with the rule. When the former users tried to convert the graph to the STAR syntax, the system's debugging messages guided them to delete the connected literal as well before proceeding with he conversion. Some participants did not notice that some literals were existing and when they tried to add them, the debugging messages informed them that the literal they were trying to add already existed, and then they proceeded with connecting the existing literal with the rule.

\textbf{Task 5b}: All participants completed this task successfully. The majority of experts, when instructed to delete a rule, they commented it out, whereas non-experts proceeded with erasing it. Some experts also used the search functionality to find the rule and then delete it.

\textbf{Task 6}: All participants completed this task successfully. They found the filtering options very easily. Some participants tried to use the filtering option before the story comprehension process was completed and they could not, since the option was available only after the completion of the process. Hence, they waited for the reading process to finish and then tried to apply the filter.

\textbf{Task 7}: All participants completed this task successfully. The majority of participants had difficulty in locating the share button. First, they searched for it on the menu bar and then at the story area. Only after careful examination of the screen they were able to locate it. Some participants browsed to the save story window and chose the ``private/public'' toggle to share the story. In most cases, participants scrolled up and down the IDE page to find the relevant control to share a story.

\textbf{Task 8}: All participants completed this task successfully. Some participants had difficulties locating how to comment on a story. They searched for the button on the menu bar and then they navigated to the public story repository to find the commenting functionality.

\textbf{Task 9}: All participants completed this task successfully. Some participants did not locate the ``Start Collaboration'' button immediately and searched for it in the public story repository.

\subsubsection{Results from the post-task questionnaire}
Results from the post-task questionnaire show that for \textbf{Task 1}, on average, participants strongly agree that the process of creating a new account \texttt{($M_E$=5.0, $M_{NE}$=4.9)\footnote{$M_{E}$ and $M_{NE}$ represent the means of the Likert scale scores given by expert and non-expert participants, respectively.}} and activating it \texttt{($M_E$=4.8, $M_{NE}$=4.9)} is easy and is the same \texttt{(for creating, $M_E$=4.6, $M_{NE}$=4.9)}, \texttt{(for activation, $M_E$=4.8, $M_{NE}$=5.0)} with that of the other systems they are using.

For \textbf{Task 2}, on average, experts agree and non-experts strongly agree that it is easy to find and start the guided tour \texttt{($M_E$=4.4, $M_{NE}$=4.8)}. On average, both experts and non-experts strongly agree that the duration of the guided tour is appropriate for learning the basics of the IDE \texttt{($M_E$=4.6, $M_{NE}$=4.5)}. In terms of feeling confident in using the IDE after the guided tour, on average, experts strongly agree that this is the case and non-experts agree as well \texttt{($M_E$=4.6, $M_{NE}$=3.9)}.

For \textbf{Task 3a}, on average, participants strongly agree that it is easy to write the story in natural language \texttt{($M_E$=5.0, $M_{NE}$=4.9)} and automatically convert it to the STAR syntax \texttt{($M_E$=5.0, $M_{NE}$=5.0)}. On average, experts agree and non-experts strongly agree that it is easy to add the background knowledge of the story using the visual editor \texttt{($M_E$=4.4, $M_{NE}$=4.5)}. Both groups strongly agree that the automatic conversion of the background knowledge in visual format to the STAR syntax is easy \texttt{($M_E$=5.0, $M_{NE}$=5.0)}. As for saving the story, participants strongly agree that it is an easy task \texttt{($M_E$=5.0, $M_{NE}$=5.0)}. Four non-experts have used the online help facility to perform this task and on average, they strongly agree that the help available from the system to perform this task is adequate \texttt{($M_{NE}$=4.8)}.

For \textbf{Task 3b}, on average, participants strongly agree that it is easy to write the story in the STAR syntax \texttt{($M_E$=5.0, $M_{NE}$=4.6)}. On average, experts agree and non-experts strongly agree that it is more efficient to write the story in natural language and then convert it to the STAR syntax than writing the story directly using the STAR syntax \texttt{($M_E$=4.4, $M_{NE}$=5.0)}. On average, participants strongly agree that it is easy to add the background knowledge in the source code editor \texttt{($M_E$=5.0, $M_{NE}$=4.9)}. One non-expert stated that this does not apply. Both groups on average, strongly agree that it is easy to convert the background knowledge from the STAR syntax to visual format \texttt{($M_E$=5.0, $M_{NE}$=5.0)}. Both experts and non-experts, on average, agree that it is easier to understand the background knowledge rules in visual format than in the STAR syntax \texttt{($M_E$=3.6, $M_{NE}$=4.3)}. Although the means of the two groups on this question appear to have a difference larger than that of other questions, further analysis revealed that this difference was not found to be statistically significant based on the Mann-Whitney test, \textit{U}$=$14.00, \textit{z}$=-$1.42, \textit{p}$>$.05, \textit{r}$=-$0.37. Participants strongly agree that it is easy to save the story. None of the participants has used the online help facility to perform this task.

For \textbf{Task 4}, on average, participants strongly agree that it is easy to find a story and load it \texttt{($M_E$=5.0, $M_{NE}$=5.0)} and that the story load window is easy to use \texttt{($M_E$=5.0, $M_{NE}$=4.9)}. On average, experts agree and non-experts strongly agree that it is easy to find how to initiate the story comprehension process \texttt{($M_E$=4.2, $M_{NE}$=4.8)}. On average, both groups strongly agree that the system provides continuous feedback on the comprehension process status \texttt{($M_E$=4.8, $M_{NE}$=4.9)}. In terms of finding the answer that the system gave to a question, on average, both experts and non-experts strongly agree that it is easy to find the answer to the question using the visual output panel \texttt{($M_E$=5.0, $M_{NE}$=4.7)}. One expert participant stated that this does not apply since he/she used only the raw output. On average, experts strongly agree and non-experts agree that it is easy to find the answer to the question using the raw output panel \texttt{($M_E$=5.0, $M_{NE}$=4.0)}. Five participants (1 expert and 4 non-experts) stated that this does not apply since they used only the visual output. On average, both experts and non-experts strongly agree that the visual output panel presents the story concepts and questions in an understandable way \texttt{($M_E$=4.8, $M_{NE}$=4.7)}. For the raw output panel, on average, experts strongly agree and non-experts agree that it presents the various story concepts and questions in an understandable way \texttt{($M_E$=5.0, $M_{NE}$=4.0)}. Five participants (1 expert and 4 non-experts) stated that this does not apply since they used only the visual output. Two participants (one from each group) used the online help facility and all participants were able to find the correct answer to the question.

For \textbf{Task 5a}, on average, experts agree and non-experts strongly agree that it is easy to add a rule using the background knowledge visual editor \texttt{($M_E$=4.4, $M_{NE}$=4.9)}. On average, both experts and non-experts strongly agree that it is easy to delete \texttt{($M_E$=4.6, $M_{NE}$=4.8)} and edit \texttt{($M_E$=5.0, $M_{NE}$=4.8)} a rule using the background knowledge visual editor. Regarding the controls available in the background knowledge visual editor, on average, experts strongly agree and non-experts agree that they are easy to use \texttt{($M_E$=4.6, $M_{NE}$=4.4)}. On average, both experts and non-experts strongly agree that it is easy to understand the functionality of the controls in the visual editor's toolbar \texttt{($M_E$=4.8, $M_{NE}$=4.6)}. Only one non-expert participant has used the online help facility. All participants were able to find the correct answer to the question.

For \textbf{Task 5b}, on average, both experts and non-experts strongly agree that it is easy to add, delete and edit a rule using the background knowledge source code editor \texttt{($M_E$=5.0, $M_{NE}$=5.0)}. Moreover, on average, both groups strongly agree that the controls available in the background knowledge source code editor's toolbar are easy to use \texttt{($M_E$=4.8, $M_{NE}$=5.0)}. On average, experts strongly agree and non-experts agree that it is easy to understand what is the functionality of the controls in the background knowledge source code editor \texttt{($M_E$=4.8, $M_{NE}$=4.9)}. One non-expert participant stated that this does not apply. In terms of what is the most efficient method to modify the background knowledge, on average, experts neither agree nor disagree that it is the visual editor, whereas non-experts agree that the visual editor is more efficient than the source code editor \texttt{($M_E$=2.8, $M_{NE}$=4.0)}. One non-expert participant stated that this does not apply. This difference between the means of the two groups was not found to be statistically significant based on the Mann-Whitney test, \textit{U}$=$10.50, \textit{z}$=-$1.65, \textit{p}$>$.05, \textit{r}$=-$0.44. None of the participants had used the online help facility. All participants but one, were able to find the correct answer to the question.

For \textbf{Task 6}, on average, both experts and non-experts strongly agree that it is easy to find and apply the filtering functionality \texttt{($M_E$=5.0, $M_{NE}$=5.0)}. Moreover, on average, participants strongly agree that the filters available can help extract information from the comprehension model \texttt{($M_E$=4.8, $M_{NE}$=4.9)}. None of the participants had used the online help facility.

For \textbf{Task 7}, on average, both experts and non-experts strongly agree that it is easy to find a demo story and load it \texttt{($M_E$=5.0, $M_{NE}$=5.0)} and that the story browser window is easy to use \texttt{($M_E$=5.0, $M_{NE}$=5.0)}. On average, both groups agree that it is easy to find how to share a story \texttt{($M_E$=4.0, $M_{NE}$=3.5)}. None of the participants had used the online help facility.

For \textbf{Task 8}, on average, experts agree and non-experts strongly agree that it is easy to find a story in the public story repository \texttt{($M_E$=4.4, $M_{NE}$=4.9)}. On average, both groups agree that it is easy to comment on a story \texttt{($M_E$=4.4, $M_{NE}$=4.4)} and strongly agree that comments added by others are clearly presented on the screen \texttt{($M_E$=4.8, $M_{NE}$=4.6)}.

For \textbf{Task 9}, on average, both experts and non-experts strongly agree that it is easy to find how to initiate the collaboration functionality \texttt{($M_E$=4.6, $M_{NE}$=4.9)}. On average, experts agree and non-experts strongly agree that the collaboration functionality could be useful for teaching logic programming \texttt{($M_E$=4.4, $M_{NE}$=4.7)}, collaboratively creating stories \texttt{($M_E$=4.4, $M_{NE}$=4.7)} and collaboratively designing knowledge \texttt{($M_E$=4.2, $M_{NE}$=4.7)}.

For all tasks, on average, participants strongly agree that the feedback messages from the system are helpful.

\subsubsection{Results from the logging functionality}
\begin{figure}
\begin{center}
\includegraphics[width=1.0\textwidth]{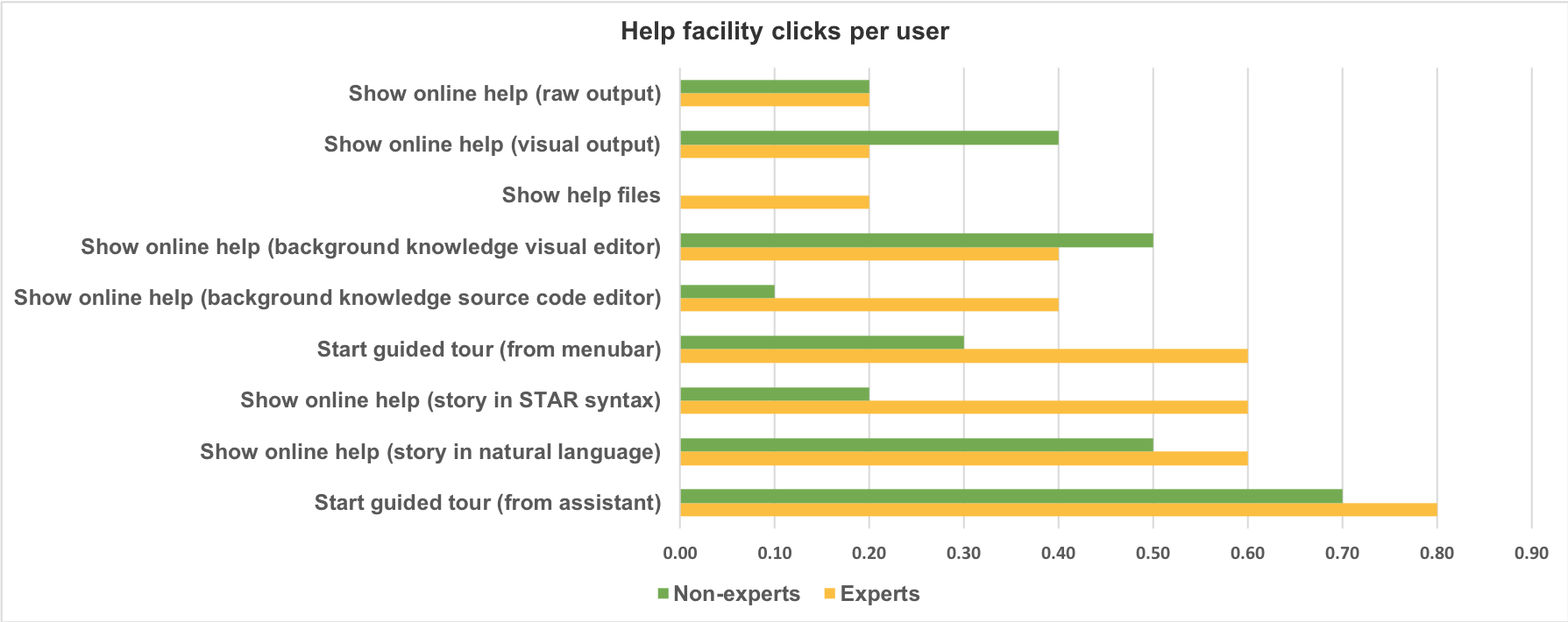}
\caption{Mean number of clicks per user for expert and non-expert participants on the help options of the IDE.}
\label{fig:evaluation_analysis:help_facility_clicks}
\end{center}
\end{figure}

\begin{figure}
\begin{center}
\includegraphics[width=1.0\textwidth]{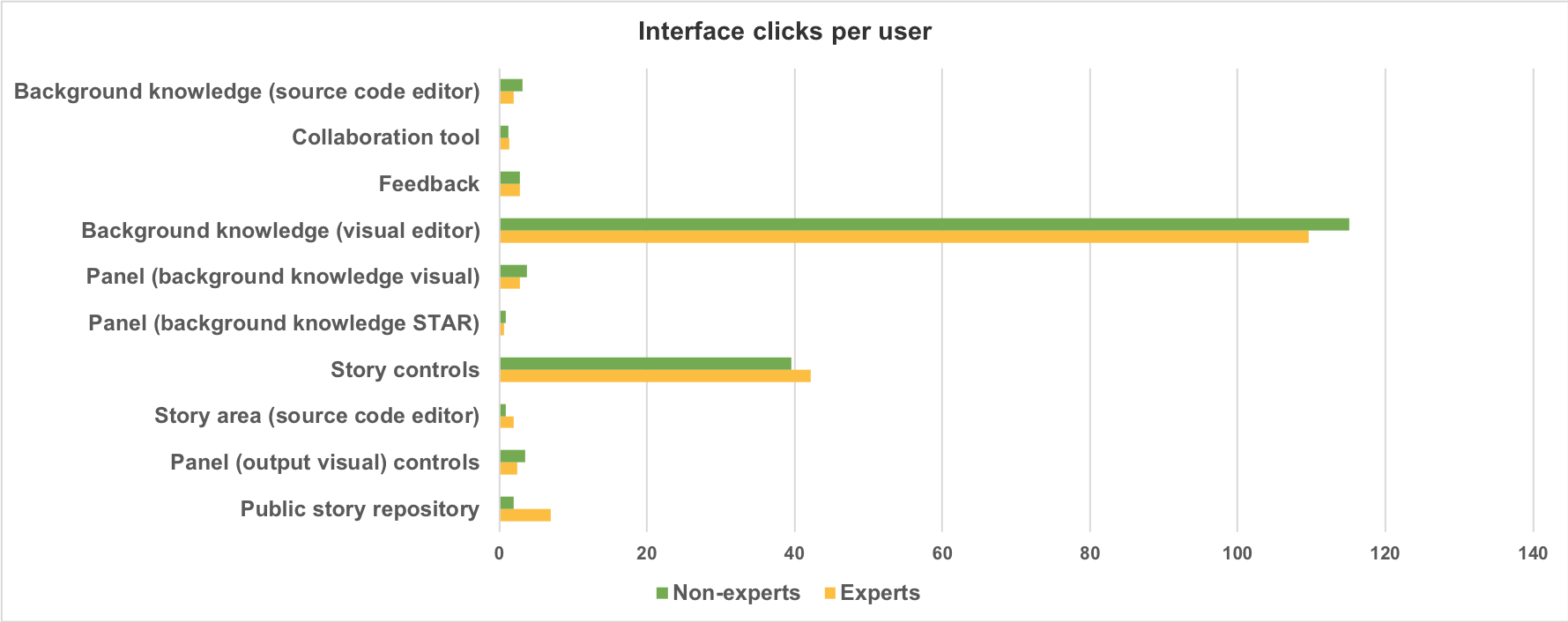}
\caption{The 10 most clickable parts of the interface per participant. The X-axis represents the mean number of clicks.}
\label{fig:evaluation_analysis:interface_clicks}
\end{center}
\end{figure}

During the experiment, all participants' interactions with the IDE were captured and stored in the database. The clicks per user for both experts and non-experts are presented in the following graphs, with a focus on the clicks on the help facilities (see Fig.~\ref{fig:evaluation_analysis:help_facility_clicks}), and on the 10 most clicked functions per user (see Fig.~\ref{fig:evaluation_analysis:interface_clicks}).

As the results show, the background knowledge visual editor is the most clickable area. This was expected since participants had to draw and edit knowledge rules using the visual editor. In general, experts and non-experts had little difference in the number of clicks per area and function.

\subsubsection{Results from the System Usability Scale (SUS) questionnaire}
Results from the System Usability Scale (SUS) standardized questionnaire show an average score of \textbf{88.33} out of \textbf{100}. The maximum score of the participants was 100, the minimum was 70 and the standard deviation was 8.5. Results are depicted in Table \ref{table:sus_all} for both groups as well as for the entire set of participants.

Compared to the SUS scores obtained from the evaluations of other systems, the Web-STAR is ranked in the top category, i.e., between ``excellent'' and ``best imaginable'' in the adjective ratings scale \cite{bangor2009determining}.

\begin{table}
\caption{Results of the System Usability Scale (SUS) standardized questionnaire}
\label{table:sus_all}
\begin{minipage}{\textwidth}
\begin{tabular}{lcccc}
\hline\hline
Group& Average score& Std\footnote{Standard deviation.}& Max\footnote{The maximum score.}& Min\footnote{The minimum score.}\\
\hline
      Experts& 90.00& 7.07& 95.00& 77.50\\
       Non-Experts& 88.25& 9.43&  100&  70\\
       \hline
       \textbf{TOTAL}& \textbf{88.83}& 8.50&  100&  70\\
             \hline\hline
    \end{tabular}
    \vspace{-2\baselineskip}
  \end{minipage}
\end{table}

\subsection{Analysis of results}
\label{subsection:results_analysis}
The evaluation process followed allowed a thorough investigation of the participants' actions, impressions, and feedback while using the IDE. The combination of a cognitive walkthrough, with questionnaires, and with close monitoring offered information that could not have obtained if using only a single method for evaluation. The diverse group of participants in this evaluation gives insights into how people from different backgrounds and prior experience in using story understanding systems and IDEs in general can benefit from the various features of the Web-STAR IDE.

After examination of the results, we report that all participants, experts and non-experts, managed to complete all the tasks. In general, participants did not have much difficulty while performing the tasks. For some tasks, like sharing a story and commenting on it, participants had some trouble finding the relevant controls since they were not in the ``expected'' area of the IDE (e.g., the menu bar).

Both experts and non-experts managed to setup an account, activate it and access the IDE in less than 2 minutes time. Participants were able to do that because the registration process is similar to that of other online systems they already have accounts on and use. They were able to start using the IDE in a very short time, by following the guided tour.

For the main task that the IDE facilitates, which is writing stories, both experts and non-experts were able to encode stories either by converting them from natural language to the STAR syntax or by writing them directly in the STAR syntax. Regarding the background knowledge, participants were able to encode it easily using the visual editor and the source code editor (even if they just had to copy the prepared story in symbolic format). All participants agree that it is easier to write the story in natural language and then convert it to the STAR syntax than writing it directly in symbolic format. Moreover, all participants were able to understand the background knowledge rules when using the visual editor and the graph representation of the background knowledge. This was clear by the answers given to the post-task questionnaire and from the time the participants took to complete the relevant tasks. This is important, since participants can use the component that best fits their working style and needs, to perform this action.

For editing the background knowledge, expert participants found the usage of the source code editor more efficient than that of the visual editor. This is to be expected, since it is presumably more time-consuming to draw a rule using the visual editor than to write it in the source code editor. For non-expert users, this was clearly not the case, since they agree that the visual editor is more efficient for changing the background knowledge. We assume that this could be because they can understand better the graph representation of the knowledge instead of the STAR syntax that they are not familiar with.

In terms of finding the answer to the questions posed, all participants were able to perform this task quite easily using either the visual or the raw output panel. Experts preferred the raw output which was enhanced with color highlighting for questions and answers and non-experts preferred the visual output with the time-line format. A number of non-expert participants chose to use only the visual output to find the answer. This is justified by the fact that the answer could be easily extracted from the time-line without the need to explore the raw output.

At this point, we observed that when a story had several scenes and a participant tried to find the answer to a question from the first scene using the raw output, he/she needed to scroll up to find it. Hence, we decided to add the option in the next version of the system to split the raw output to scenes, so that this user burden can be avoided, by making it easier to browse each scene from the raw output. Both groups benefited from the time-line format since it was easier to understand the various concepts of each story, apply filters on them, and find answers to questions.

The social and collaboration features of the IDE were also simple to use. Both groups were able to share a story or add comments to a story in the public story repository in a very short time. We observed that a number of participants had a problem spotting the relevant controls, since they were not located in the expected area. Hence, we decided to add a menu option that groups all these controls and buttons together for easy access in the next version of the system. Participants also found the collaboration feature very useful, since they confirmed that it could be useful for teaching logic programming, collaboratively creating stories, and collaboratively designing knowledge.

Results from the SUS standardized questionnaire dictate that the Web-STAR IDE is a friendly, easy to use, and easy to learn IDE. This evaluation led to some minor changes in the IDE to enhance user experience and productivity.

\section{Conclusion and future work}
\label{section:discussion}
We have presented Web-STAR, a platform built to facilitate the interaction of users with the STAR system for story comprehension. We have discussed the various features of the platform through examples, and have argued that the platform is designed to appeal to both expert and non-expert users. The argument is supported, in particular, by the visualization that Web-STAR offers for the background knowledge that is used during story comprehension, and for the output of the story comprehension process. A comprehensive evaluation of the usability of the platform has supported that the platform is, indeed, friendly, easy to use, and easy to learn.

The platform is currently used for educational purposes, helping students and researchers engage with the problem of automated story understanding. More than 40 users have registered so far, and have contributed more than 70 stories. Moreover, the platform has received more than 1350 web service calls for processing STAR domain files. 
The platform was also used in the Robot Trainer Game \cite{Rodosthenous2016}, a crowdsourcing game with a purpose to gather commonsense knowledge; in this context, knowledge contributed by users was processed in real-time to determine its sufficiency to answer story questions. The demonstrable ease of use of the Web-STAR platform, and its online and visual environment, makes it a prime candidate for use by domain experts in law, history, or literature, who may wish to comprehend text in the form of narratives.

Future versions of the platform will aim to refine its interface and extend its functionality. In terms of the latter, we are considering the addition of the option to import and process relevant background knowledge from existing knowledge bases like Conceptnet \cite{Speer2013}, YAGO \cite{Mahdisoltani2015}, NELL \cite{NELL-aaai15} and the OpenCyc project \cite{Lenat1995}; or directly Games With A purpose \cite{VonAhn2008} or other crowdsourcing platforms like Amazon Mechanical Turk \cite{doi:10.1177/1745691610393980}; or even from machine learning algorithms that produce rule-based knowledge bases \cite{Michael_2009_ReadingBetweenTheLines,Michael_2016_CognitiveMechanisms,Michael_2017_TheAdviceTaker2.0}. The component that converts natural language stories into the STAR syntax could be further extended, by incorporating systems that extract knowledge from natural language \cite{2016tkde}, and identify the temporal ordering of events \cite{UzZaman2013}.

It is our hope that this work will serve as a basis for establishing a story-sharing and story-processing community, towards the advancement of work in automated story understanding through symbolic knowledge and reasoning.

\section*{Acknowledgments}
The authors would like to thank: Adamos Koumis, for his collaboration in the development of the component that converts natural language to the STAR syntax; Elektra Kypridemou, for her help in the preparation of the evaluation methodology; and the anonymous reviewers, for their valuable comments and suggestions.

\section*{Appendix A - Cognitive Walkthrough Tasks}
\label{section:appendix1}
\includepdf[pages={1-}]{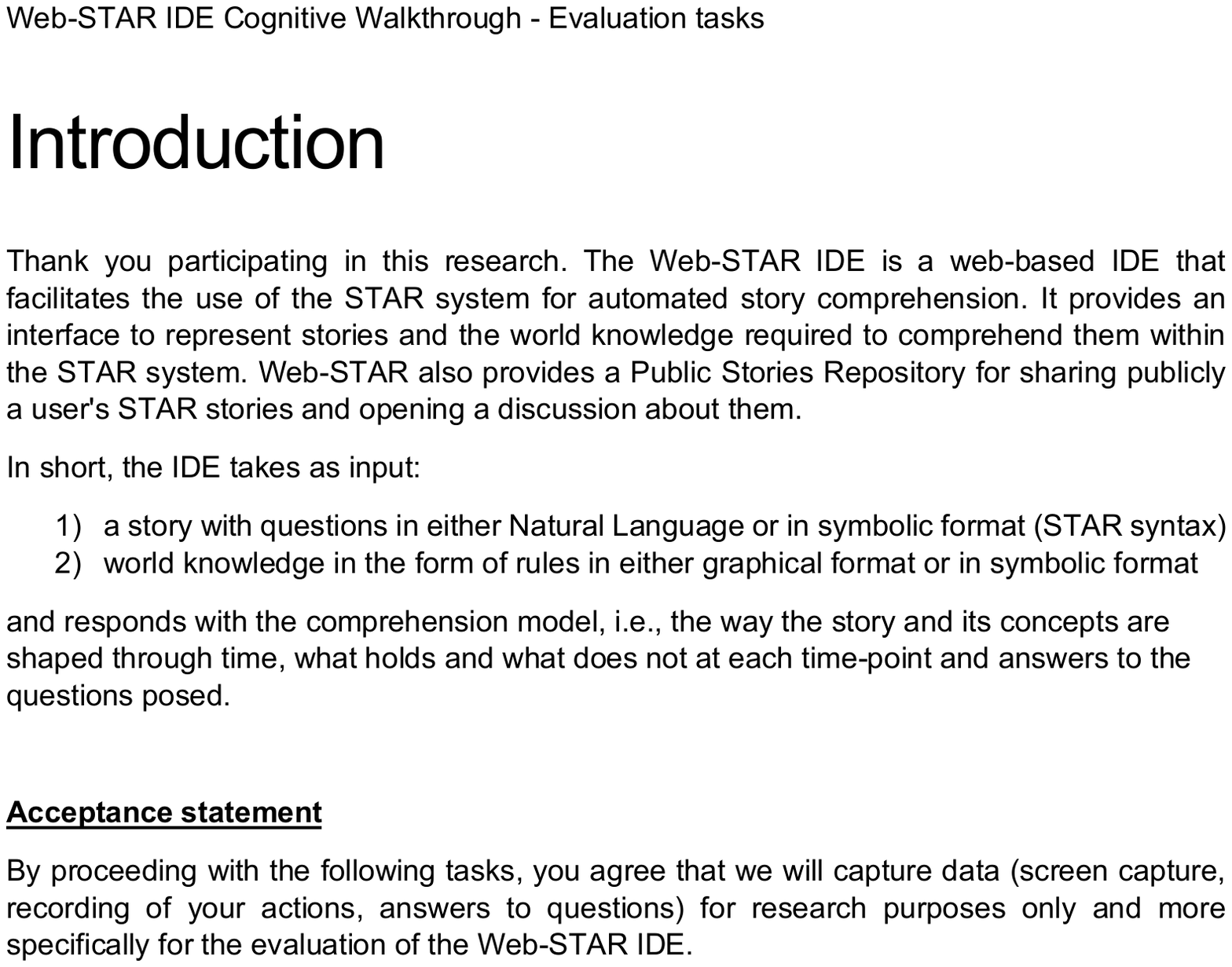}

\section*{Appendix B - Web-STAR IDE Evaluation Questionnaire}
\label{section:appendix2}


\section*{Supplementary material}

# Web-STAR evaluation Questionnaire

## Introduction

Thank you participating in this research. The Web-STAR IDE is a web-based IDE that facilitates the use of the STAR system for automated story comprehension. It provides an interface to represent stories and the world knowledge required to comprehend them within the STAR system. Web-STAR also provides a Public Stories Repository for sharing publicly a user's STAR stories and opening a discussion about them.

In short, the IDE takes as input:

1. a story with questions in either Natural Language or in symbolic format (STAR syntax)
2. world knowledge in the form of rules in either graphical format or in symbolic format

and responds with the comprehension model, i.e., the way the story and its concepts are shaped through time, what holds and what does not at each timepoint and answers to the questions posed.


**Acceptance statement**

By proceeding with the following tasks, you agree that we will capture data (screen capture, recording of your actions, answers to questions) for research purposes only and more specifically for the evaluation of the Web-STAR IDE.

There are 41 questions in this survey


## General (demographics)
## []Please type your experiment ID *

Please write your answer here:

<br>

## []Please select your gender *

Please choose **only one** of the following:

○ Female

○ Male

## []Please select your age group *

Choose one of the following answers

Please choose **only one** of the following:

○ 18-24 years old

○ 25-34 years old

○ 35-44 years old

○ 45-54 years old

○ 55-64 years old

○ 65-74 years old

○ 75 years or older

# []What is the highest degree or level of school you have completed? *

Choose one of the following answers

Please choose **only one** of the following:

○ Less than a high school diploma

○ High school degree or equivalent (e.g. GED)

○ Some college, no degree

○ Associate degree (e.g. AA, AS)

○ Bachelor's degree (e.g. BA, BS)

○ Master's degree (e.g. MA, MS, MEd)

○ Professional degree (e.g. MD, DDS, DVM)

○ Doctorate (e.g. PhD, EdD)

○

If you're currently enrolled in school, please indicate the highest degree you have *received*

# []What is your current employment status? *

Choose one of the following answers

Please choose **only one** of the following:

○ Employed full time (40 or more hours per week)

○ Employed part time (up to 39 hours per week)

○ Unemployed and currently looking for work

○ Unemployed and not currently looking for work

○ Student

○ Retired

○ Homemaker

○ Self-employed

○ Unable to work

○

# []Your degree is relevant to: *

Check all that apply

Please choose **all** that apply:

☐ Computer Science

☐ Psychology

☐ Philosophy

☐ Storytelling or Narratology

☐ Linguistics

- ☐ Law
- ☐ Other: _____

# General (development specific)

## []Have you ever used an Integrated Development Environment (IDE) before for software development or programming? *

Please choose **only one** of the following:

○ Yes

○ No

## []Please specify what applies for each of the following IDEs: *

**Only answer this question if the following conditions are met:**
Answer was 'Yes' at question '7 [B01]' (Have you ever used an Integrated Development Environment (IDE) before for software development or programming?)

Please choose the appropriate response for each item:

| | 1: I have never heard about it | 2: I have heard about it but I have never used it | 3: I have heard about it | 4: I have used it before | 5: It is on of the IDEs I mostly use |
|---|---|---|---|---|---|
| Microsoft Visual Studio | ○ | ○ | ○ | ○ | ○ |
| NetBeans | ○ | ○ | ○ | ○ | ○ |
| Eclipse | ○ | ○ | ○ | ○ | ○ |
| Cloud9 | ○ | ○ | ○ | ○ | ○ |
| Codiad | ○ | ○ | ○ | ○ | ○ |
| ICEcoder | ○ | ○ | ○ | ○ | ○ |
| Codeanywhere | ○ | ○ | ○ | ○ | ○ |
| Eclipse Che | ○ | ○ | ○ | ○ | ○ |

## []Please specify what applies for each of the following programming languages: *

Please choose the appropriate response for each item:

| | 1: I have never heard about it | 2: I have heard about it but I have never used it | 3: I have heard about it | 4: I have used it before | 5: It is one of the programming languages I use frequently |
|---|---|---|---|---|---|
| c | ○ | ○ | ○ | ○ | ○ |
| c++ | ○ | ○ | ○ | ○ | ○ |
| JAVA | ○ | ○ | ○ | ○ | ○ |
| javascript | ○ | ○ | ○ | ○ | ○ |
| PHP | ○ | ○ | ○ | ○ | ○ |
| Python | ○ | ○ | ○ | ○ | ○ |
| Perl | ○ | ○ | ○ | ○ | ○ |
| Prolog | ○ | ○ | ○ | ○ | ○ |
| Lisp | ○ | ○ | ○ | ○ | ○ |

# []Are you familiar with the notion of automated story understanding by machines? *

Please choose **only one** of the following:

○ Yes

○ No

# []Have you ever used a story understanding system? *

Please choose **only one** of the following:

○ Yes

○ No

# []Have you ever used the STAR story understanding system? *

**Only answer this question if the following conditions are met:**
Answer was 'Yes' at question '11 [B05]' (Have you ever used a story understanding system?)

Please choose **only one** of the following:

○ Yes

○ No

# Task 1

## []Please answer the degree at which you agree or disagree with the following statements: *

Please choose the appropriate response for each item:

|  | Strongly disagree | Disagree | Neither agree nor disagree | Agree | Strongly agree | Does not apply |
|---|---|---|---|---|---|---|
| The process of creating a new account is easy. | ○ | ○ | ○ | ○ | ○ | ○ |
| The process of creating a new account is the same as with the other systems I use. | ○ | ○ | ○ | ○ | ○ | ○ |
| The process of activating the new account is easy. | ○ | ○ | ○ | ○ | ○ | ○ |
| The process of activating the new account is the same as with the other systems I use. | ○ | ○ | ○ | ○ | ○ | ○ |
| The feedback messages from the system while performing the task are helpful. | ○ | ○ | ○ | ○ | ○ | ○ |

# Task 2

[]Please answer the degree at which you agree or disagree with the following statements: *

Please choose the appropriate response for each item:

|  | Strongly disagree | Disagree | Neither agree nor disagree | Agree | Strongly agree | Does not apply |
|---|---|---|---|---|---|---|
| It is easy to find and start the guided tour. | ○ | ○ | ○ | ○ | ○ | ○ |
| The duration of the guided tour is appropriate for learning the basics of the IDE. | ○ | ○ | ○ | ○ | ○ | ○ |
| After completing the guided tour, I feel confident in using the IDE. | ○ | ○ | ○ | ○ | ○ | ○ |

# Task 3a

## []Please answer the degree at which you agree or disagree with the following statements: *

Please choose the appropriate response for each item:

|  | Strongly disagree | Disagree | Neither agree nor disagree | Agree | Strongly agree | Does not apply |
|---|---|---|---|---|---|---|
| It is easy to write the story in natural language. | ○ | ○ | ○ | ○ | ○ | ○ |
| The automatic conversion of the story from natural language to STAR syntax is easy . | ○ | ○ | ○ | ○ | ○ | ○ |
| It is easy to add the background knowledge of the story using the visual editor. | ○ | ○ | ○ | ○ | ○ | ○ |
| The automatic conversion of the background knowledge in visual format to STAR syntax is easy. | ○ | ○ | ○ | ○ | ○ | ○ |
| It is easy to save the story. | ○ | ○ | ○ | ○ | ○ | ○ |
| The feedback messages from the system while performing the task are helpful. | ○ | ○ | ○ | ○ | ○ | ○ |

## []I have used the online help facility to perform this task. *

Please choose **only one** of the following:

○ Yes

○ No

## []Please answer the degree at which you agree or disagree with the following statements: *

**Only answer this question if the following conditions are met:**
Answer was 'Yes' at question '16 [TSK03a2]' (I have used the online help facility to perform this task.)

Please choose the appropriate response for each item:

|  | Strongly disagree | Disagree | Neither agree nor disagree | Agree | Strongly agree | Does not apply |
|---|---|---|---|---|---|---|
| The help available from the system to perform this task is adequate. | ○ | ○ | ○ | ○ | ○ | ○ |

# Task 3b

## []Please answer the degree at which you agree or disagree with the following statements: *

Please choose the appropriate response for each item:

| | Strongly disagree | Disagree | Neither agree nor disagree | Agree | Strongly agree | Does not apply |
|---|---|---|---|---|---|---|
| It is easy to write the story in STAR syntax. | ○ | ○ | ○ | ○ | ○ | ○ |
| It is more efficient to write the story in natural language and then convert it to STAR syntax than writing the story directly using the STAR syntax. | ○ | ○ | ○ | ○ | ○ | ○ |
| It is easy to add the background knowledge in the source code editor | ○ | ○ | ○ | ○ | ○ | ○ |
| It is easy to convert the background knowledge from STAR syntax to visual format. | ○ | ○ | ○ | ○ | ○ | ○ |
| It is easier to understand the background knowledge rules in visual format than in STAR syntax. | ○ | ○ | ○ | ○ | ○ | ○ |
| It is easy to save the story. | ○ | ○ | ○ | ○ | ○ | ○ |
| The feedback messages from the system while performing the task are helpful. | ○ | ○ | ○ | ○ | ○ | ○ |

## []I have used the online help facility to perform this task. *

Please choose **only one** of the following:

○ Yes

○ No

## []Please answer the degree at which you agree or

# disagree with the following statements: *

**Only answer this question if the following conditions are met:**
Answer was 'Yes' at question '19 [TSK03b2]' (I have used the online help facility to perform this task.)

Please choose the appropriate response for each item:

|  | Strongly disagree | Disagree | Neither agree nor disagree | Agree | Strongly agree | Does not apply |
|---|---|---|---|---|---|---|
| The help available from the system to perform this task is adequate. | ○ | ○ | ○ | ○ | ○ | ○ |

# Task 4

[]
What answer <u>the system gave</u> to the multiple choice question posed in the story at session 2?

*

Choose one of the following answers

Please choose **only one** of the following:

○ accepted choice: [penguin at 9], accepted choice: [bird at 9], rejected choice: [flying at 9]

○ rejected choice: [penguin at 9], accepted choice: [bird at 9], accepted choice: [flying at 9]

## []Please answer the degree at which you agree or disagree with the following statements: *

Please choose the appropriate response for each item:

| | Strongly disagree | Disagree | Neither agree nor disagree | Agree | Strongly agree | Does not apply |
|---|---|---|---|---|---|---|
| It is easy to find a story and load it. | ○ | ○ | ○ | ○ | ○ | ○ |
| The Load Story window is easy to use. | ○ | ○ | ○ | ○ | ○ | ○ |
| It is easy to find how to initiate the story comprehension process. | ○ | ○ | ○ | ○ | ○ | ○ |
| The system provides constant feedback on the comprehension process status. | ○ | ○ | ○ | ○ | ○ | ○ |
| It is easy to find the answer to the question using the visual output panel (left panel). | ○ | ○ | ○ | ○ | ○ | ○ |
| It is easy to find the answer to the question using the raw output panel (right panel). | ○ | ○ | ○ | ○ | ○ | ○ |

|  | Strongly disagree | Disagree | Neither agree nor disagree | Agree | Strongly agree | Does not apply |
|---|---|---|---|---|---|---|
| The visual output panel presents the story concepts and questions in an understandable way. | ○ | ○ | ○ | ○ | ○ | ○ |
| The raw output panel presents the various story concepts and questions in an understandable way. | ○ | ○ | ○ | ○ | ○ | ○ |
| The feedback messages from the system while performing the task are helpful. | ○ | ○ | ○ | ○ | ○ | ○ |

## []I have used the online help facility to perform this task. *

Please choose **only one** of the following:

○ Yes

○ No

## []Please answer the degree at which you agree or disagree with the following statements: *

**Only answer this question if the following conditions are met:**
Answer was 'Yes' at question '23 [TSK042]' (I have used the online help facility to perform this task.)

Please choose the appropriate response for each item:

|  | Strongly disagree | Disagree | Neither agree nor disagree | Agree | Strongly agree | Does not apply |
|---|---|---|---|---|---|---|
| The help available from the system to perform this task is adequate. | ○ | ○ | ○ | ○ | ○ | ○ |

# Task 5a

[]
## What answer <u>the system gave</u> to the multiple choice question posed in the story at session 2?

*

Choose one of the following answers

Please choose **only one** of the following:

○ rejected choice: [on_fire(the_house) at 1]

○ accepted choice: [on_fire(the_house) at 1]

## []Please answer the degree at which you agree or disagree with the following statements: *

Please choose the appropriate response for each item:

|  | Strongly disagree | Disagree | Neither agree nor disagree | Agree | Strongly agree | Does not apply |
|---|---|---|---|---|---|---|
| It is easy to add a rule using the Background Knowledge in Visual Format editor. | ○ | ○ | ○ | ○ | ○ | ○ |
| It is easy to delete a rule using the Background Knowledge in Visual Format editor. | ○ | ○ | ○ | ○ | ○ | ○ |
| It is easy to edit a rule using the Background Knowledge in Visual Format editor. | ○ | ○ | ○ | ○ | ○ | ○ |
| The controls available in the Background Knowledge in Visual Format editor are easy to use. | ○ | ○ | ○ | ○ | ○ | ○ |

|  | Strongly disagree | Disagree | Neither agree nor disagree | Agree | Strongly agree | Does not apply |
|---|---|---|---|---|---|---|
| It is easy to understand what is the functionality of the controls in the Background Knowledge in Visual Format editor toolbar. | ○ | ○ | ○ | ○ | ○ | ○ |
| The feedback messages from the system while performing the task are helpful. | ○ | ○ | ○ | ○ | ○ | ○ |

# []I have used the online help facility to perform this task. *

Please choose **only one** of the following:

○ Yes

○ No

# []Please answer the degree at which you agree or disagree with the following statements: *

**Only answer this question if the following conditions are met:**
Answer was 'Yes' at question '27 [TSK05a2]' (I have used the online help facility to perform this task.)

Please choose the appropriate response for each item:

|  | Strongly disagree | Disagree | Neither agree nor disagree | Agree | Strongly agree | Does not apply |
|---|---|---|---|---|---|---|
| The help available from the system to perform this task is adequate. | ○ | ○ | ○ | ○ | ○ | ○ |

# Task 5b

[]
What answer <u>the system gave</u> to the multiple choice question posed in the story at session 1?

*

Choose one of the following answers

Please choose **only one** of the following:

○ accepted choice: [on_fire(the_house) at 1]

○ rejected choice: [on_fire(the_house) at 1]

## []Please answer the degree at which you agree or disagree with the following statements: *

Please choose the appropriate response for each item:

|  | Strongly disagree | Disagree | Neither agree nor disagree | Agree | Strongly agree | Does not apply |
|---|---|---|---|---|---|---|
| It is easy to add a rule using the Background Knowledge source code editor. | ○ | ○ | ○ | ○ | ○ | ○ |
| It is easy to delete a rule using the Background Knowledge source code editor. | ○ | ○ | ○ | ○ | ○ | ○ |
| It is easy to edit a rule using the Background Knowledge source code editor. | ○ | ○ | ○ | ○ | ○ | ○ |
| The controls available in the Background Knowledge source code editor toolbar are easy to use. | ○ | ○ | ○ | ○ | ○ | ○ |
| It is easy to understand what is the functionality of the controls in the Background Knowledge source code editor. | ○ | ○ | ○ | ○ | ○ | ○ |

| | Strongly disagree | Disagree | Neither agree nor disagree | Agree | Strongly agree | Does not apply |
|---|---|---|---|---|---|---|
| It is more efficient to use the Background Knowledge in Visual Format editor to modify the background knowledge than the Background Knowledge source code editor. | ○ | ○ | ○ | ○ | ○ | ○ |
| The feedback messages from the system while performing the task are helpful. | ○ | ○ | ○ | ○ | ○ | ○ |

# []I have used the online help facility to perform this task. *

Please choose **only one** of the following:

○ Yes

○ No

# []Please answer the degree at which you agree or disagree with the following statements: *

**Only answer this question if the following conditions are met:**
Answer was 'Yes' at question '31 [TSK05b2]' (I have used the online help facility to perform this task.)

Please choose the appropriate response for each item:

| | Strongly disagree | Disagree | Neither agree nor disagree | Agree | Strongly agree | Does not apply |
|---|---|---|---|---|---|---|
| The help available from the system to perform this task is adequate. | ○ | ○ | ○ | ○ | ○ | ○ |

# Task 6

## []Please answer the degree at which you agree or disagree with the following statements: *

Please choose the appropriate response for each item:

|  | Strongly disagree | Disagree | Neither agree nor disagree | Agree | Strongly agree | Does not apply |
|---|---|---|---|---|---|---|
| It is easy to find the filtering functionality. | ○ | ○ | ○ | ○ | ○ | ○ |
| It is easy to apply the filter on the comprehension model. | ○ | ○ | ○ | ○ | ○ | ○ |
| The filters available can help me extract information from the comprehension model. | ○ | ○ | ○ | ○ | ○ | ○ |
| The feedback messages from the system while performing the task are helpful. | ○ | ○ | ○ | ○ | ○ | ○ |

## []I have used the online help facility to perform this task. *

Please choose **only one** of the following:

○ Yes

○ No

## []Please answer the degree at which you agree or disagree with the following statements: *

**Only answer this question if the following conditions are met:**
Answer was 'Yes' at question '34 [TSK062]' (I have used the online help facility to perform this task.)

Please choose the appropriate response for each item:

|  | Strongly disagree | Disagree | Neither agree nor disagree | Agree | Strongly agree | Does not apply |
|---|---|---|---|---|---|---|
| The help available from the system to perform this task is adequate. | ○ | ○ | ○ | ○ | ○ | ○ |

# Task 7

## []Please answer the degree at which you agree or disagree with the following statements: *

Please choose the appropriate response for each item:

| | Strongly disagree | Disagree | Neither agree nor disagree | Agree | Strongly agree | Does not apply |
|---|---|---|---|---|---|---|
| It is easy to find a demo story and load it. | ○ | ○ | ○ | ○ | ○ | ○ |
| The story browser window is easy to use. | ○ | ○ | ○ | ○ | ○ | ○ |
| It is easy to find how to share a story. | ○ | ○ | ○ | ○ | ○ | ○ |
| The feedback messages from the system while performing the task are helpful. | ○ | ○ | ○ | ○ | ○ | ○ |

## []I have used the online help facility to perform this task. *

Please choose **only one** of the following:

○ Yes

○ No

## []Please answer the degree at which you agree or disagree with the following statements: *

**Only answer this question if the following conditions are met:**
Answer was 'Yes' at question '37 [TSK072]' (I have used the online help facility to perform this task.)

Please choose the appropriate response for each item:

| | Strongly disagree | Disagree | Neither agree nor disagree | Agree | Strongly agree | Does not apply |
|---|---|---|---|---|---|---|
| The help available from the system to perform this task is adequate. | ○ | ○ | ○ | ○ | ○ | ○ |

# Task 8

[]Please answer the degree at which you agree or disagree with the following statements: *

Please choose the appropriate response for each item:

|  | Strongly disagree | Disagree | Neither agree nor disagree | Agree | Strongly agree | Does not apply |
|---|---|---|---|---|---|---|
| It is easy to find a story in the public story repository. | ○ | ○ | ○ | ○ | ○ | ○ |
| It is easy to comment on a story. | ○ | ○ | ○ | ○ | ○ | ○ |
| Comments added by others are clearly presented on the screen. | ○ | ○ | ○ | ○ | ○ | ○ |

# Task 9

## []Please answer the degree at which you agree or disagree with the following statements: *

Please choose the appropriate response for each item:

| | Strongly disagree | Disagree | Neither agree nor disagree | Agree | Strongly agree | Does not apply |
|---|---|---|---|---|---|---|
| It is easy to find how to initiate the collaboration functionality. | ○ | ○ | ○ | ○ | ○ | ○ |
| The collaboration functionality could be useful for teaching logic programming. | ○ | ○ | ○ | ○ | ○ | ○ |
| The collaboration functionality is useful for collaborative creating of stories. | ○ | ○ | ○ | ○ | ○ | ○ |
| The collaboration functionality is useful for collaborative designing of knowledge. | ○ | ○ | ○ | ○ | ○ | ○ |
| The feedback messages from the system to perform the task are helpful. | ○ | ○ | ○ | ○ | ○ | ○ |

# System Usability Scale

[]Please answer the degree at which you agree or disagree with the following statements: *

Please choose the appropriate response for each item:

|  | Strongly disagree | Disagree | Neither agree nor disagree | Agree | Strongly agree |
|---|---|---|---|---|---|
| I think that I would like to use the Web-STAR IDE frequently. | ○ | ○ | ○ | ○ | ○ |
| I found the Web-STAR IDE unnecessarily complex. | ○ | ○ | ○ | ○ | ○ |
| I thought the Web-STAR IDE was easy to use. | ○ | ○ | ○ | ○ | ○ |
| I think that I would need the support of a technical person to be able to use the Web-STAR IDE. | ○ | ○ | ○ | ○ | ○ |
| I found the various functions in the Web-STAR IDE were well integrated. | ○ | ○ | ○ | ○ | ○ |
| I thought there was too much inconsistency in the Web-STAR IDE. | ○ | ○ | ○ | ○ | ○ |
| I would imagine that most people would learn to use the Web-STAR IDE very quickly. | ○ | ○ | ○ | ○ | ○ |
| I found the Web-STAR IDE very awkward to use. | ○ | ○ | ○ | ○ | ○ |
| I felt very confident using the Web-STAR IDE. | ○ | ○ | ○ | ○ | ○ |
| I needed to learn a lot of things before I could get going with the Web-STAR IDE. | ○ | ○ | ○ | ○ | ○ |

Thank you for participating.


Submit your survey.
Thank you for completing this survey.



\bibliography{webstar}

\end{document}